\documentclass[11pt]{article}
\usepackage{subfigure}
\usepackage{hyperref,graphicx}

\begin{document}

%%%% Article title to be placed here
\title{New approximations, and policy implications, from a delayed dynamic model of a fast pandemic}

\author{C. P. Vyasarayani$^{1}$ \and Anindya Chatterjee$^{2}$}

%%%%%%%%% Insert author address here
\date{$^{1}$Mechanical and Aerospace Engineering, Indian Institute of Technology Hyderabad, Sangareddy, 502285, India.
Email: vcprakash@mae.iith.ac.in\\
	$^{2}$Mechanical Engineering, Indian Institute of Technology Kanpur, Kanpur, 208016, India. Email: anindya100@gmail.com, anindya@iitk.ac.in\\
\bigskip
\bigskip \today}

\maketitle

%\subject{Applied mathematics, mathematical modelling, differential equations}

%\keywords{Multiple scales, long-wave solution, epidemic, COVID-19, social distancing}

%\corres{C. P. Vyasarayani\\
%\email{vcprakash@mae.iith.ac.in}}

%%%% Abstract text to be placed here %%%%%%%%%%%%
\begin{abstract}
We study an SEIQR (Susceptible-Exposed-Infectious-Quarantined-Recovered) model for an infectious disease, with time delays for latency and an asymptomatic phase. For fast pandemics where nobody has prior immunity and everyone has immunity after recovery, the SEIQR model decouples into two nonlinear delay differential equations (DDEs) with five parameters. One parameter is set to unity by scaling time. The subcase of perfect quarantining and zero self-recovery before quarantine, with two free parameters, is examined first. The method of multiple scales yields a hyperbolic tangent solution; and a long-wave approximation yields a first order ordinary differential equation (ODE). With imperfect quarantining and nonzero self-recovery, the long-wave approximation is a second order ODE. These three approximations each capture the full outbreak, from infinitesimal initiation to final saturation. Low-dimensional dynamics in the DDEs is demonstrated using a six state non-delayed reduced order model obtained by Galerkin projection. Numerical solutions from the reduced order model match the DDE over a range of parameter choices and initial conditions. Finally, stability analysis and numerics show how correctly executed time-varying social distancing, within the present model, can cut the number of affected people by almost half. Alternatively, faster detection followed by near-certain quarantining can potentially be even more effective.\\

\end{abstract}
%%%%%%%%%%%%%%%%%%%%%%%%%%%

\maketitle
\section{Introduction}
This work is partially motivated by the global pandemic of COVID-19. Understanding the dynamics of infectious
diseases in a population can help in developing strategies
to mitigate the spread~\cite{vynnycky2010introduction,capasso2008mathematical}.
This paper presents new mathematical approximations and an asymptotic solution for a specific dynamic model for such infectious diseases.
Some policy implications are discussed as well.

Mathematical models for the spread of disease have almost a century old history. In their seminal paper, Kermack and McKendrick~\cite{kermack1927contribution} proposed a three-state model (popularly known as SIR) governing the evolution of susceptible (S), infected (I), and recovered (R) populations. In their model, the recovered population is assumed to have developed immunity against the infection. The model contains two free parameters, one for infection rate and one for recovery rate. The SIR model is widely used to predict the number of infected people in closed populations. The model has an analytical solution. Over time, the SIR model~\cite{brauer2012mathematical} has been modified to study infections where the recovered population can be reinfected (as with the common cold) and is known as the two-state SIS model. In the classic endemic  model~\cite{hethcote1989three}, for diseases that are active over $10$-$20$ years, information of new births and deaths are included. In another variant of the SIR model known as the four-state MSIR~\cite{hethcote2009basic} model, passive immunity inherited by newborns from their mothers is included: for example, newborn babies can be immune to measles for some time after their birth, but become susceptible later on.
Other modifications have considered the effect of a carrier population~\cite{hethcote2009basic}, which never recovers from the disease but is asymptomatic (relevant to, e.g., tuberculosis). Such people can again suffer from the disease later, or continue to infect others while remaining asymptomatic. In SEIR~\cite{hethcote2000mathematics}, a four-state model, one of the states (E) represents the exposed population, infected but non-infectious. In the SEIQR model~\cite{gerberry2009seiqr}, yet another state, representing a quarantined population, is added to the SEIR model. All the models discussed above, including SIR, SIS, MSIR, SEIR, and SEIQR are governed by nonlinear differential equations. More complicated partial differential equation models that include the effect of the age structure~\cite{brauer2012mathematical} of the population and vaccination history are also available~\cite{hethcote2000mathematics}. 

The models mentioned so far need not include time delays. However,
the incubation, asymptomatic, and symptomatic phases of a disease can be incorporated as time delays in mathematical models. Including such delays in the differential equation models make them delay differential equations (DDEs), also known as retarded functional differential equations. Researchers have used DDEs~\cite{young2019consequences,bocharov2000numerical} to model the spread of infections like Zika~\cite{rakkiyappan2019fractional}, HIV~\cite{nelson2002mathematical}, influenza~\cite{alexander2008delay}, and Hepatitis B~\cite{gourley2008dynamics}. Recently, a scalar DDE with one delay (time of recovery) based on a logistic model was used to study the spread of COVID-19 in Italy~\cite{Dell2020}. Some policy implications based on parameter studies and simulations were reported, but analytical progress on the delayed equation was limited.

We observe that the abovementioned models, with a few states, can be formally developed from underlying network models.
Network epidemiological models with a large number of states~\cite{zuzek2015epidemic, morita2016six,hasegawa2017efficiency,strona2018rapid,coelho2008epigrasss,keeling2005networks} follow fine details of the spatial and temporal spread of an infection.  Their average or overall behavior can be described using what we might call lumped, compartmental, or continuum models.

In a recent article~\cite{Singh2020age}, a network of compartmental models, with each model representing an age group was used to study the COVID-19 outbreak in India. Other works that have recently appeared, which study the progression of COVID-19 in different parts of the world, include~\cite{Crokidakis2020, Savi2020, Roques2020SIR, Radulescu2020SEIR}. These works primarily deal with parameter identification, numerical simulations, and prediction of the number of cases with time, along with some policy implications.

Even with lumped models, if various  effects~\cite{brauer2012mathematical} like incubation times, natural birth and death rates, prior immunity, and carrier states are included, then analytical solutions are usually unavailable. However, sometimes difficult problems can be solved approximately using asymptotic methods~\cite{hinchperturbation,Kevorkian1995} or related methods, and this paper presents such approximations for a slightly simplified case that is relevant to a fast-spreading pandemic.

In recent work that is directly related to our paper~\cite{young2019consequences}, a five-state SEIQR system with delays has been developed as a continuum limit from a network model under quite general conditions. That system has then been examined as a generic model for infectious diseases. The model, even after simplification, has multiple parameters and coupled states, and so analytical progress is difficult. However, several interesting limits, steady states, and stability criteria have been presented~\cite{young2019consequences}, along with supporting numerical results and policy implications.

Here we take up a simplified version of that model~\cite{young2019consequences}. Our simplification is only that we ignore the possibility of some past sufferers of the disease eventually losing their immunity, and becoming vulnerable to infection all over again. With this simplification, significant new analytical approximations and insights are possible, and constitute the contribution of this paper. We note that Young {\em et al.}~\cite{young2019consequences} say the following (their state of $(1,0,0,0,0)$ is an uninfected population):
\begin{quote}
	``The fact
	that orbits starting from near $(1, 0, 0, 0, 0)$ will approach an endemic equilibrium is difficult to prove; this part of
	our prediction is supported by numerical simulations.''
\end{quote}
In this context, we will present a new asymptotic multiple-scales solution for weak growth in a special case, and two informal long-wave approximations for moderate growth, all three solutions describing the complete evolution from infinitesimal infection to final saturation. These new approximations provide useful new analytical support and understanding that is not included within~\cite{young2019consequences}. We will also discuss transient solutions and the role of initial conditions, in the context of a time-varying infection rate, using a six-state Galerkin approximation based reduced-order model obtained from the original
delayed equations. Since social distancing affects the infection rate, we will demonstrate how, within the present model, properly executed social distancing can cut the total number of affected people by almost one half. A stronger case for early detection and quarantine will emerge simultaneously.

\section{Mathematical Model}
\begin{figure}[h!]
	\centerline{{\includegraphics[width=0.8\textwidth]{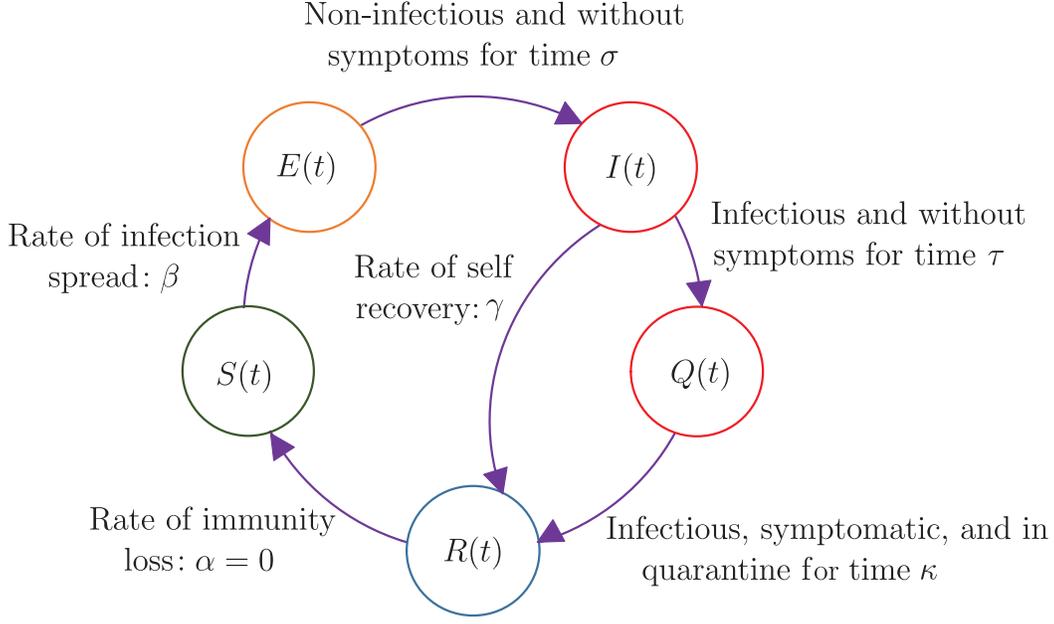}}}
	\caption{The SEIQR model with delays. Healthy individuals $S(t)$ are infected with rate constant $\beta$. Infected individuals $E(t)$ remain asymptomatic and non-infectious for a time duration $\sigma$. Subsequently, these individuals become infectious and enter population $I(t)$, but remain asymptomatic for a time duration $\tau$. Upon showing symptoms, they enter population $Q(t)$ and are quarantined with probability $p$ for a time $\kappa$, beyond which they infect nobody.  Some infectious asymptomatic individuals may become non-infectious on their own, with rate $\gamma$. After quarantine, the cured population $R(t)$ could in principle lose immunity at a small rate $\alpha$, but we take $\alpha = 0$ for a fast-spreading pandemic.}
	\label{Figure1}
\end{figure}

As mentioned above, our mathematical model is essentially that of Young {\em et al.}~\cite{young2019consequences}, with one of their small parameters set to zero. Figure \ref{Figure1}, adapted from their paper, shows the lumped model which is itself obtained by them from an underlying network model. In the model there are five subpopulations (actually population fractions) that add up to unity.
The general governing equations~\cite{young2019consequences} are:
\begin{eqnarray}
\label{Sdot1}
\dot{S}(t)&=&-\tilde \beta m S(t)I(t)+\alpha R(t)\\
\label{Edot1}
\dot{E}(t)&=&\tilde \beta m\left[S(t)I(t)-S(t-\sigma)I(t-\sigma)\right]\\
\label{Idot1}
\dot{I}(t)&=&\tilde \beta m S(t-\sigma)I(t-\sigma)-\gamma I(t)-\tilde \beta m pe^{-\gamma\tau}S(t-\sigma-\tau)I(t-\sigma-\tau)\\
\label{Qdot1}
\dot{Q}(t)&=&\tilde \beta m pe^{-\gamma\tau}\left[S(t-\sigma-\tau)I(t-\sigma-\tau)-S(t-\sigma-\tau-\kappa)I(t-\sigma-\kappa)\right]\\
\label{Rdot1}
\dot{R}(t)&=&-\alpha R(t)+\gamma I(t)+\tilde \beta m pe^{-\gamma\tau} S(t-\sigma-\tau-\kappa)I(t-\sigma-\tau-\kappa)
\end{eqnarray}
In the above equations, the free parameters are interpreted as follows: $\tilde \beta$ is the infection rate, $m$ is the density of contacts, $\gamma$ is the self-recovery rate, $p$ is the probability of identifying and isolating an infected individual, and $\alpha$ is the rate of immunity loss. We have not introduced any simplifications of our own so far.

We note, first, that $E$ and $Q$ in equations (\ref{Edot1}) and (\ref{Qdot1}) are influenced by $S$ and $I$ along with their delayed values, but $E$ and $Q$ do not themselves influence $S$, $I$ and $R$. In other words, $E$ and $Q$ are slave variables, and we henceforth ignore them.

In the three equations that remain, we can clearly absorb $m$ into $\tilde \beta$ or equivalently, write
\begin{equation}
\label{betascale} \tilde \beta m=\beta.
\end{equation}
If social distancing is practiced, then we expect $m$ to decrease and thus $\beta$ to be lower, even though $\tilde \beta$ may remain the same. For this reason, in the later portions of this paper, we will consider time-varying $\beta$ while holding other parameters constant.

For a fast-spreading pandemic, we assume $\alpha = 0$ for simplicity, which makes $R$ a slave as well, and we need to only retain the equations for $S$ and $I$. Finally, by choice of units of time, we can let $\sigma = 1$. This is equivalent to nondimensionalizing $\tau$ which has units of time, as well as $\gamma$ and $\beta$ which have units of 1/time. Our equations now are
\begin{eqnarray}
\label{Sdot2a}
\dot{S}(t)&=&-\beta(t) S(t)I(t)\\
\label{Idot2a}
\dot{I}(t)&=&\beta(t-1) S(t-1)I(t-1)-\gamma I(t)-\beta(t-1-\tau)  pe^{-\gamma\tau}S(t-1-\tau)I(t-1-\tau)
\end{eqnarray}
In the above, if $\beta$ was constant rather than time-varying, then its $t$-dependence would be dropped. Note that, if $\beta$ varies with time, its variation can be considered externally specified and not a part of the solution. Equations~(\ref{Sdot2a}) and
(\ref{Idot2a}) make intuitive sense in a lumped-variable setting as follows. Equation~(\ref{Sdot2a}) says that the instantaneous rate of new infections is proportional to how infectious the disease is ($\tilde \beta$), how much people are meeting each other ($m(t)$), how many uninfected people there are ($S(t)$) and how many infectious people are out in public ($I(t)$). Equation~(\ref{Idot2a}) says that the rate of change in the number of infectious people is equal to previously infected people just exiting the latency phase and entering the infectious phase, minus the rate at which people are recovering on their own, minus also the rate at which people displaying symptoms are being put into quarantine (these quarantined people are slightly diminished in number due to self-recovery, which is good; and due to some people not being quarantined, which is a system inefficiency).

We are interested in near-unity initial conditions for $S(-\infty)=1$ that lead to growth of the infection and eventual saturation. In particular, the net damage done by the disease is represented by $1 - S(\infty).$

From now onward, until the start of section \ref{tvb}, we will consider $\beta$ to be a constant parameter.
For clarity, therefore, we write the governing equations out with constant $\beta$,
\begin{eqnarray}
\label{Sdot2}
\dot{S}(t)&=&-\beta S(t)I(t)\\
\label{Idot2}
\dot{I}(t)&=&\beta S(t-1)I(t-1)-\gamma I(t)-\beta  pe^{-\gamma\tau}S(t-1-\tau)I(t-1-\tau)
\end{eqnarray}
Let
\begin{equation}
P(t) = \int_{-\infty}^t I(\zeta) \, d \zeta,
\label{IntTrans}
\end{equation}
where we are interested in the asymptotic initial condition $P(-\infty) = 0$ as the limiting case of a tiny level of initial infection. Then equation~(\ref{Sdot2}) yields
\begin{equation}
S(t) = e^{-\beta P(t)}
\label{SsolP}
\end{equation}
which incorporates the initial condition of interest, namely $S(-\infty) = 1.$ Inserting $S(t)$ from equation~(\ref{SsolP}) into equation~(\ref{Idot2}), we obtain
\begin{equation}
\ddot P(t) = \beta  e^{-\beta P(t-1)} \dot P(t-1) - p e^{-\gamma \tau} \beta e^{-\beta P(t-1-\tau)} \dot P(t-1-\tau) 
-\gamma \dot P(t).
\label{Pddot}
\end{equation}
Integrating both sides with respect to time, and by defining 
\begin{equation} \label{pbareq} \bar{p}=p e^{-\gamma \tau},
\end{equation}
we obtain
\begin{equation}
\dot P(t) =  \bar{p}e^{-\beta P(t-1-\tau)}-e^{-\beta P(t-1)}-\gamma P(t)+1-\bar{p},
\label{Pdotnon}
\end{equation}
where $1-\bar{p}$ is an integration constant chosen to match initial conditions at $-\infty$.
Thus, for constant $\beta$ and with the approximation of $\alpha = 0$ for a fast-spreading pandemic, equations  (\ref{Sdot1}) through (\ref{Rdot1}) effectively reduce to the single nonlinear delay differential equation shown in  (\ref{Pdotnon}).

It may be noted that $\dot P = 0$, i.e., $P$ equals a constant, is allowed by equation~(\ref{Pdotnon}) for $P$ that satisfy
\begin{equation}
\label{equilval} (1 - \bar p) \left ( 1 - e^{-\beta P} \right ) -\gamma P = 0.
\end{equation}
Equation~(\ref{equilval}) is satisfied by $P=0$ for all parameter values. Additionally, for $\gamma > 0$,
it has a single strictly positive root if
\begin{equation}
\frac{\beta}{\gamma}\left( 1- \bar p \right) > 1.
\label{reduce}
\end{equation}
If $\gamma = 0$ (i.e., there is no self-recovery), and $0 \le \bar p < 1$ (i.e., not everybody is quarantined), then for $\beta > 0$, $P=0$ is the only equilibrium solution. This means if $P$ increases from zero, it can grow without bound and $S(\infty) = 0$, i.e., everybody in the population gets infected. The case of $\beta = 0$ is not interesting because the infection does not spread. Finally, if $\gamma = 0$ and $\bar p =1$, then equation~(\ref{equilval}) is identically satisfied for every constant
$P$.

Thus, we conclude that a simple yet interesting situation within equation~(\ref{Pdotnon}) occurs when $p=1$ (all infected people display symptoms and are quarantined) and $\gamma = 0$ (there is no recovery without displaying symptoms). We will first study this restricted case in some detail, because some analytical progress is possible that provides useful insights.

\section{A simple subcase: $p=1$ and $\gamma = 0$}
\label{restc}
For $p=1$ and $\gamma = 0$, we have from equation~(\ref{Pdotnon}),
\begin{equation}
\dot P(t) =  e^{-\beta P(t-1-\tau)} - e^{-\beta P(t-1)}
\label{Pdotnondim}
\end{equation}
As mentioned above, any constant $P$ is an equilibrium, though possibly an unstable one.

\subsection{Linear Stability Analysis for Small $P$}
In the initial stages $P(t)$ is small, and equation~(\ref{Pdotnondim}) can be linearized to
\begin{equation}
\dot P(t) =  \beta P(t-1)- \beta P(t-1-\tau)
\label{PdotLin}
\end{equation}
Equation~(\ref{PdotLin}) has infinitely many characteristic roots, and has oscillatory solutions which we must disallow because their decreasing portions require negative $I(t)$. However, monotonic solutions exist as well, and we will examine them.
The characteristic equation of equation~(\ref{PdotLin}) is
\begin{equation}
\lambda = \beta e^{-\lambda} \left ( 1 - e^{-\lambda \tau} \right).
\label{Pchar}
\end{equation}
Among the infinitely many roots of the above equation, those with nonnegative real parts are of main interest because they lead to growth of $P(t)$ from initial tiny values. We note that if the real part of $\lambda$ is assumed nonnegative, then the magnitude of the right hand side is bounded by $ 2\beta$. This means, for infinitesimal $\beta$, any right half plane roots of equation~(\ref{Pchar}) are infinitesimal as well. We also note that $\lambda = 0$ is a root regardless of $\beta$.

For non-infinitesimal $\beta$, a criterion for {\em real} roots in the right half plane is easily found because $\lambda = 0$ is a root of equation~(\ref{Pchar}) and the right hand side first increases and then decreases to zero as $\lambda \rightarrow \infty$. These conditions imply that a nonzero positive root is assured if the slope of the right hand side at $\lambda = 0$ exceeds unity. That condition is
\begin{equation}
\beta \tau > 1.
\label{sat0}
\end{equation}
Equation~(\ref{sat0}) suggests that the contagion may take hold if $\beta \tau > 1$, and perhaps not if $\beta \tau < 1$. Note that $\tau$ is fixed by biology but $\beta$ can be lowered by practicing social distancing, which will be discussed towards the end of the paper. 

Incidentally, in Young {\em et al.}~\cite{young2019consequences}, the instability condition including $p \le 1$ and $\gamma > 0$ is given as equation~(\ref{reduce}).
For $p=1$ and $\gamma\rightarrow0$, equation~(\ref{reduce}) easily reduces to equation~(\ref{sat0}).

\subsection{Initial Numerical Observations}

For initial insight, we consider some numerical solutions of equation~(\ref{Pdotnondim}) obtained using Matlab's built-in solver {\tt dde23} with specified error tolerances of $10^{-7}$ or better. The initial function used was
\begin{equation}
P(t) = a \left(1+\frac{t}{1+\tau}\right),\, t\le 0,
\label{Hist}
\end{equation}
with $a$ small and positive, and mentioned in the figure captions.
Once $P(t)$ is obtained (numerically or, later, analytically), we can calculate $S(t)$ by using equation~(\ref{SsolP}). Results below will therefore be presented only in terms of the original variable $S(t)$, since the impact of the disease is reflected by $1-S(\infty)$.
\begin{figure}[htpb!]
	\begin{center}
		\subfigure[]{\includegraphics[width=0.48\textwidth]{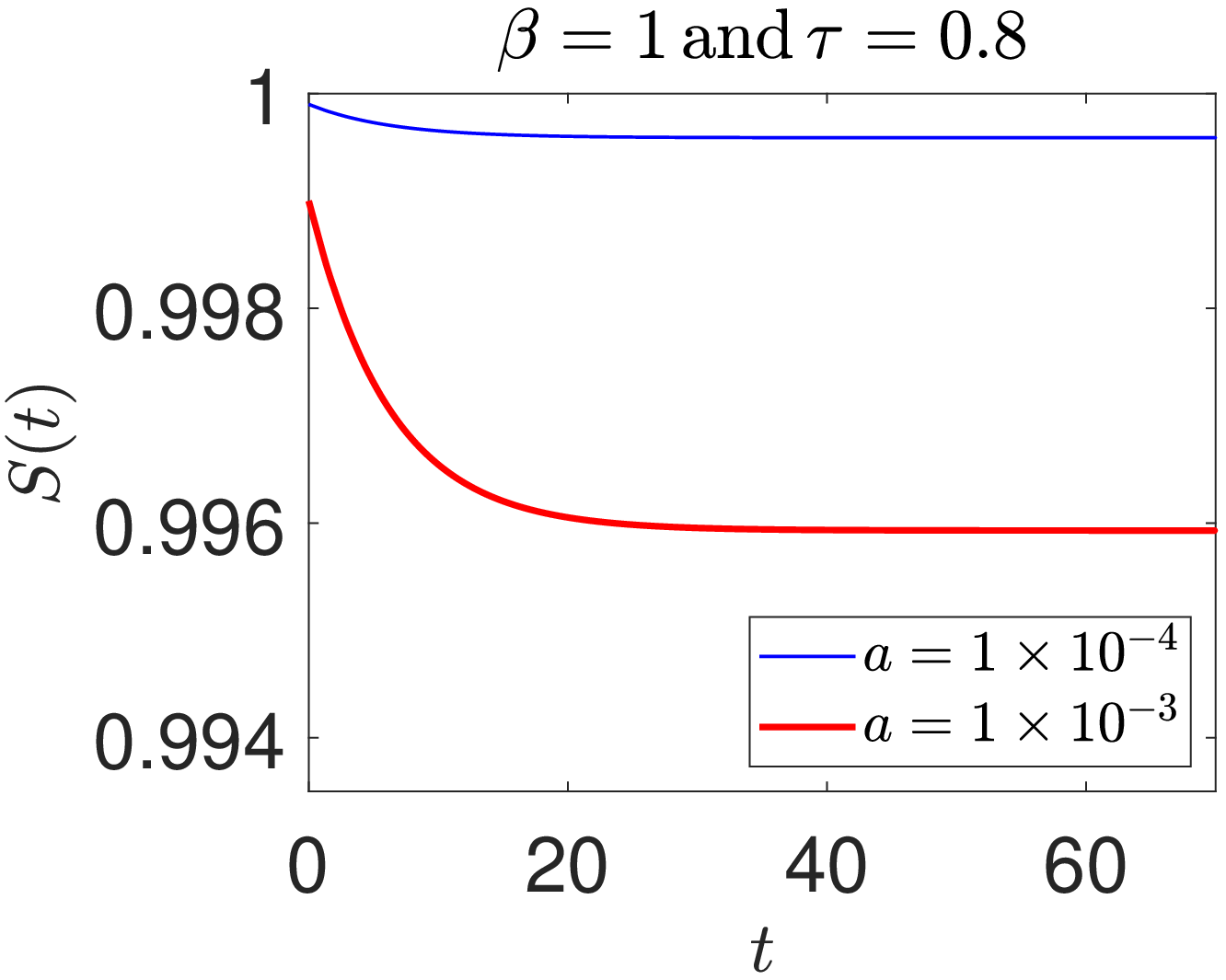}\label{Figure2a}}
		\subfigure[]{\includegraphics[width=0.48\textwidth]{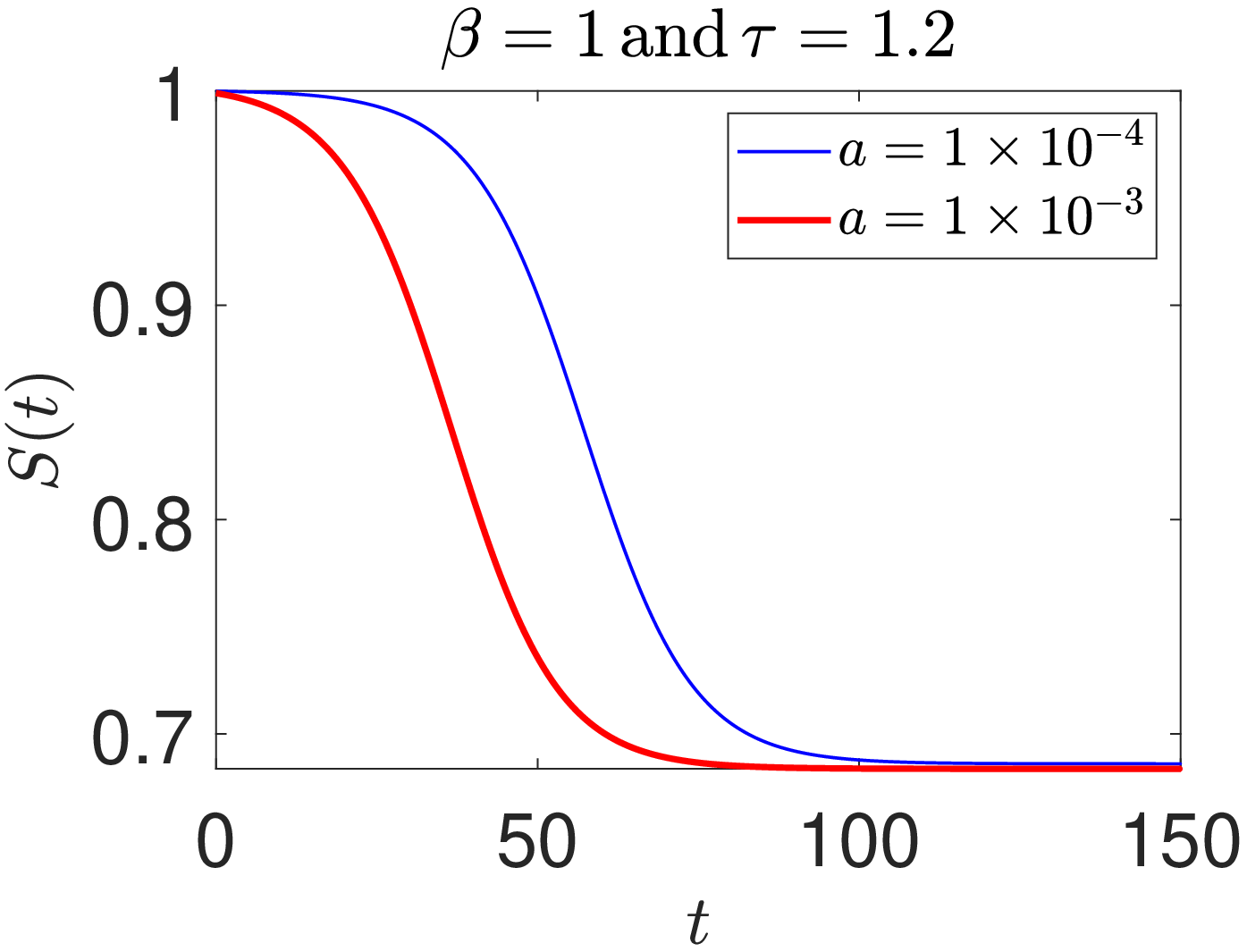}\label{Figure2b}}
	\end{center}
	\caption{Two solutions each for (a) $\beta \tau = 0.8 < 1$ and (b) $\beta \tau = 1.2 > 1$.  The initial function used  for integrating equation~(\ref{Pdotnondim}) was $P(t) = a \left(1+\frac{t}{1+\tau}\right),\, t\le 0$. The two solutions in (b) are relatively time-shifted because the underlying system is autonomous, and one solution grows from smaller initial values; see main text for further discussion.}
	\label{Figure2}
\end{figure}

Two solutions for $\beta = 1$ and $\tau=0.8$ are shown in figure~\ref{Figure2a}. It is seen that not much decay occurs; and the solution in each case saturates at a value proportional to the magnitude of the initial function used. This is linear behavior (recall also the discussion following equation~(\ref{sat0})).

In contrast, two solutions for $\beta = 1$ and $\tau=1.2$ are shown in figure~\ref{Figure2b}. Numerical solutions for two different initial functions show that the rapidly spreading phase of the disease and the saturation value of $S$ are essentially identical, independent of the initial function.
It is important to note that the relative time-shift between these two solutions is inconsequential. The underlying dynamical system is autonomous, and the use of asymptotic initial conditions at $-\infty$ leaves a free parameter in the solution that allows time-translations (this will be explicitly clear in the analytical approximations later in the paper). The solution that starts from smaller initial values takes a little longer to climb to larger values, resulting in the time-shift.

\begin{figure}[htpb!]
	\begin{center}
		\subfigure[]{\includegraphics[width=0.48\textwidth]{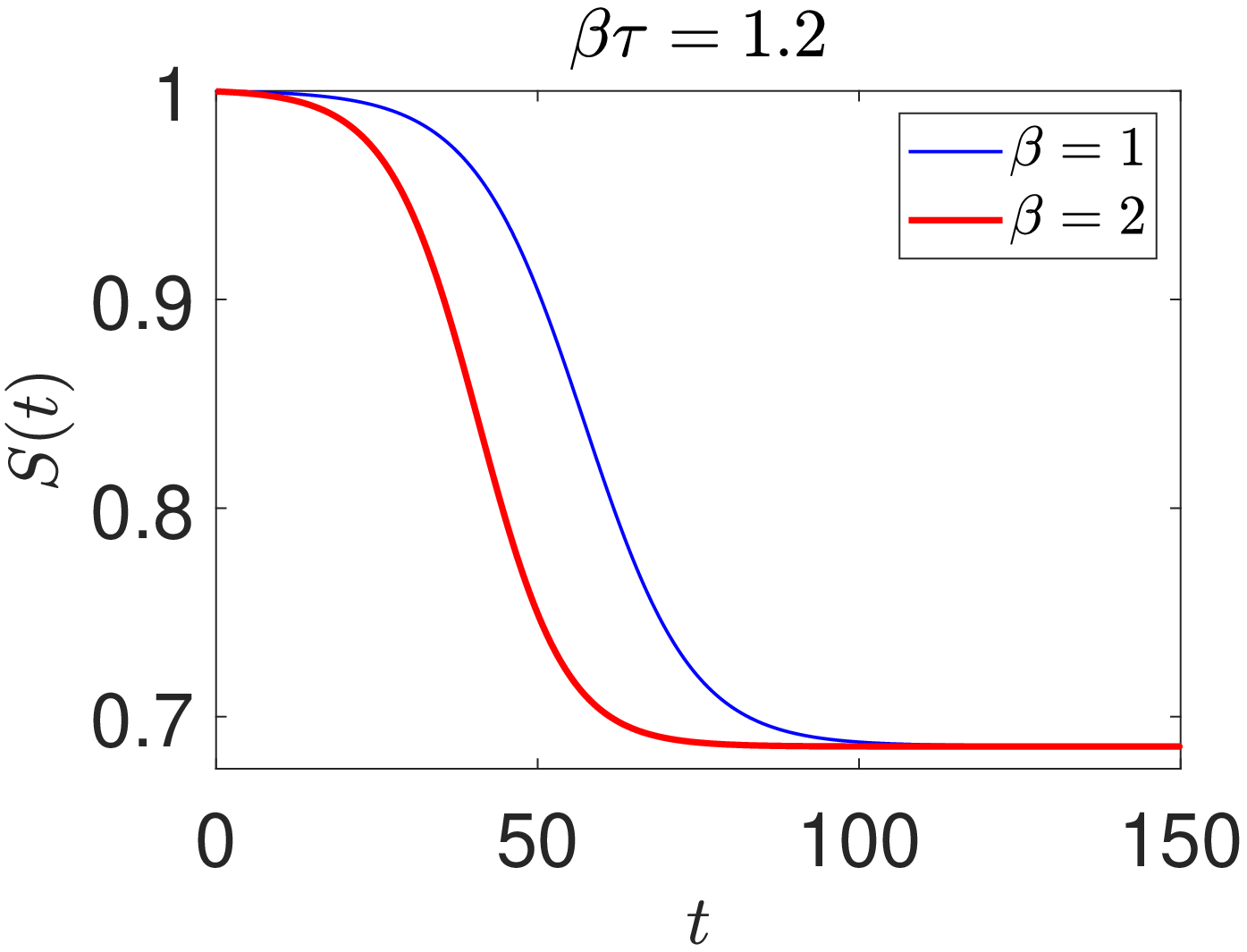}\label{Figure3a}}
		\subfigure[]{\includegraphics[width=0.48\textwidth]{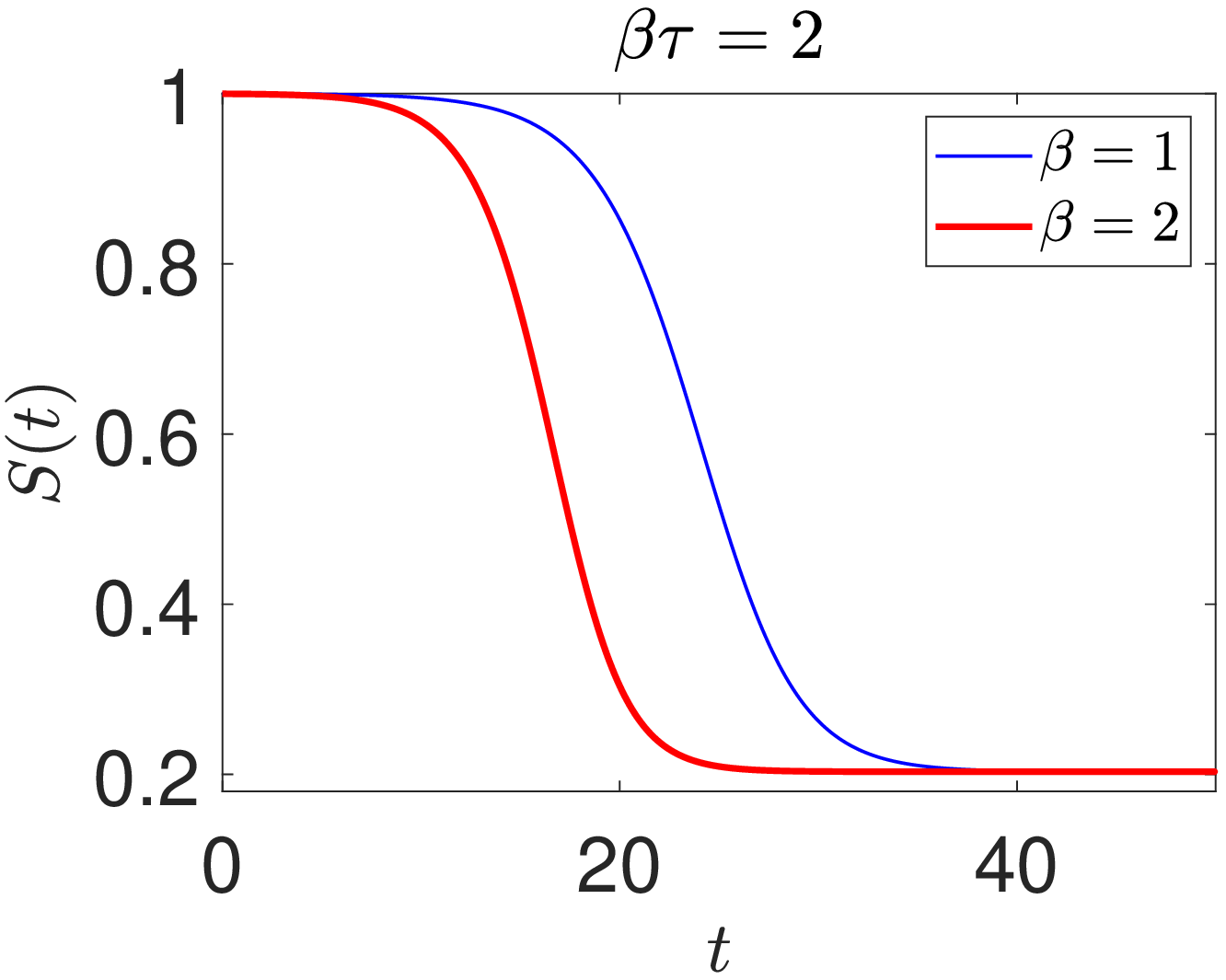}\label{Figure3b}}
	\end{center}
	\caption{Two solutions for (a) $\beta \tau = 1.2 > 1$ and (b) $\beta \tau = 2 > 1$. The parameters $\beta$ and $\tau$  are individually varied.  The initial function used  for integrating equation~(\ref{Pdotnondim}) was $P(t) = 1\times 10^{-4} \left(1+\frac{t}{1+\tau}\right),\, t\le 0$.}
	\label{Figure3}
\end{figure}

Next, two solutions for $\beta = 1$ and $\beta = 2$, but with $\beta \tau = 1.2$ in each case, and with identical initial functions ($a = 0.0001$) are shown in figure~\ref{Figure3a}.
The limiting or saturation value $S(\infty)$, which is the population fraction that remains unaffected, appears independent of $\beta$ for $\beta \tau$ held fixed. Similar behavior is seen in the two curves in figure~\ref{Figure3b} for $\beta \tau = 2$ and identical initial conditions. Again, $S(\infty)$ appears independent of $\beta$ for $\beta \tau$ held fixed.

\subsection{Saturation Value of $P$}
Prompted by numerics, let
\begin{equation}
P(\infty) = \frac{C}{\beta}
\label{Satura}
\end{equation}
for some $C$ to be determined. 
In the solution of interest, $I(t)$ starts from infinitesimal values, and each individual who is a part of $S(t)$ stays in the infectious state for exactly $\tau$ units of time. Since the total number of individuals thus affected is exactly $1-S(\infty)$, we can see that for the solution of interest\footnote{%
	An analogy may help explain this trick. Imagine a room where a finite but large number of people, say $M$ people,
	enter over a long period of time and at a variable rate.
	The number of people in the room, $N(t)$, is an arbitrary nonnegative function of time.
	Each person stays in the room for exactly $\tau$ units of time, and then leaves.
	Clearly, $\int_{-\infty}^{\infty} N(t) dt = M \tau$.}
\begin{equation}
P(\infty) = \left ( 1 - S(\infty) \right ) \tau.
\end{equation}
From equation~(\ref{Satura}), we obtain
\begin{equation}
\frac{C}{1-e^{-C}} = \beta \tau
\label{Sat5}
\end{equation}
Equation~(\ref{Sat5}) needs to be solved numerically, but two limits are clear. As $\beta\tau \rightarrow 1^+$, we have $C \rightarrow 0$. A simple calculation shows
\begin{equation}
C \approx 2 ( \beta \tau -1 ). 
\label{Cap}
\end{equation}
The other limit is for $\beta \tau \gg 1$, where $C \rightarrow \beta \tau$.

Numerically, for $\beta \tau = 1.2$, we find $C=0.376$ or $S(\infty) = e^{-0.376} \approx 0.687$, which matches figure~\ref{Figure2a}; and for $\beta \tau =2$, we find
$C=1.594$ or $S(\infty) = e^{-1.594} \approx 0.203$, which matches figure~\ref{Figure2b}. The difference between even $\beta \tau = 1.2$ and $\beta \tau =2$, though both are unstable, is large in terms of consequences for the population.

\subsection{Maximum Value of Stable $S(\infty)$}
\label{minstab}
The above results indicate that if $\beta \tau > 1$, then a solution that grows asymptotically from a zero value at minus infinity
saturates at a value given by equation~(\ref{Cap}). However, all $P$ values are equilibrium values: it is therefore interesting to ask what the minimum value of $P$ is for which the equilibrium is stable. We can find this by considering the corresponding limiting
steady value of $S$ to be $S^*$, writing
equation~(\ref{Idot2})  for $p=1$ and $\gamma = 0$ as
$$\dot{I}(t) = \beta S^* I(t-1) -\beta   S^*I(t-1-\tau),$$
and concluding that the required stability condition is (by adapting equation~(\ref{sat0}))
\begin{equation}
\label{Smax} \beta \tau S^* < 1.
\end{equation}

For $\beta \tau$ slightly exceeding unity, referring to equations~(\ref{Satura}) and (\ref{Cap}), the upper limit of
$$S^* = \frac{1}{\beta \tau}$$
implied by equation~(\ref{Smax}) corresponds to about half as many people being infected as would be if the asymptotic solution for constant $\beta$ was allowed to run its course all the way from initiation to saturation. This situation will be clearly seen
in the multiple scales solution for constant $\beta$ and weak growth starting from zero at $t = -\infty$. We now turn to that solution.

\subsection{Multiple Scales Solution for Weak Growth}
\begin{figure}[htpb!]
	\begin{center}
		\subfigure[]{\includegraphics[width=0.48\textwidth]{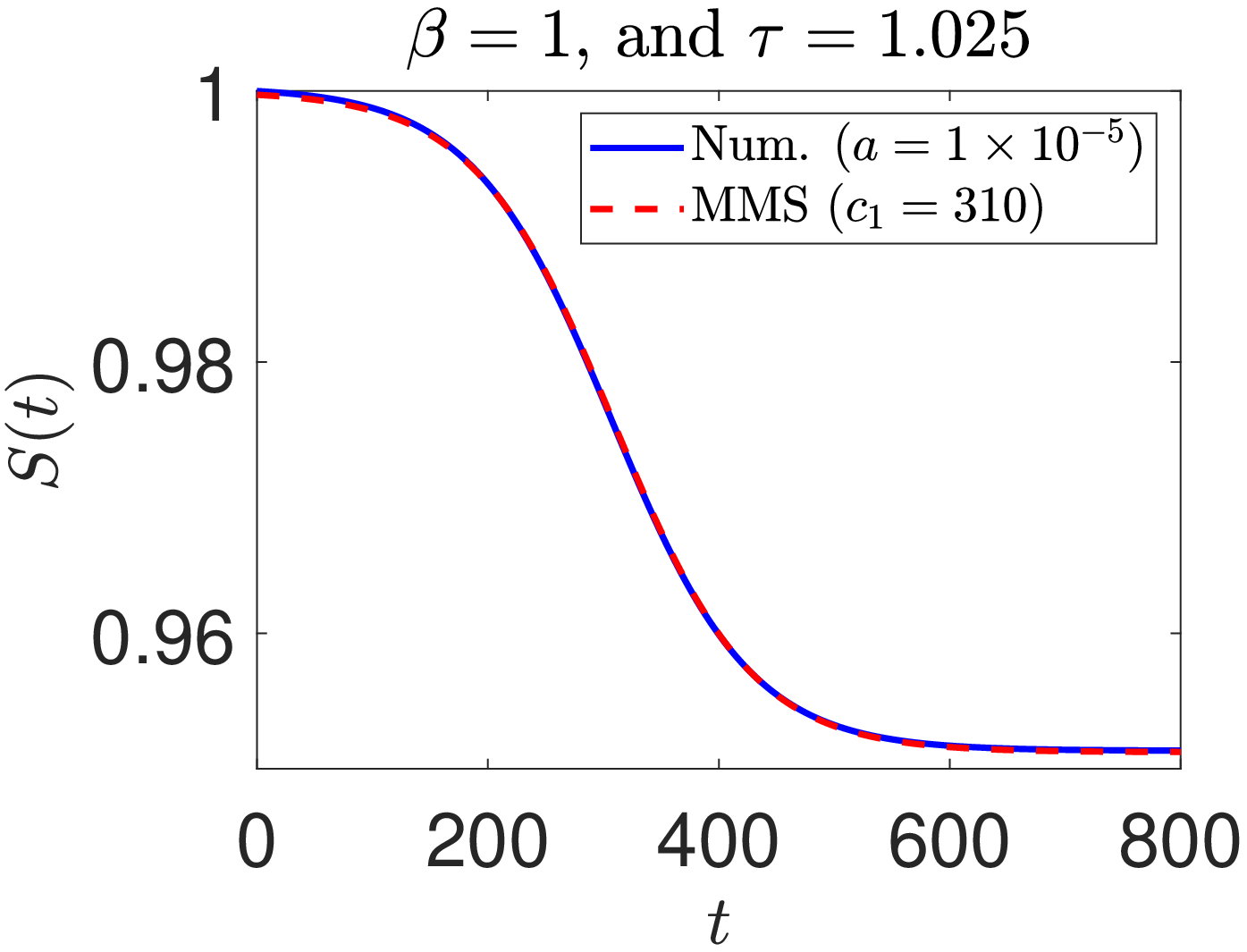}\label{Figure4a}}
		\subfigure[]{\includegraphics[width=0.48\textwidth]{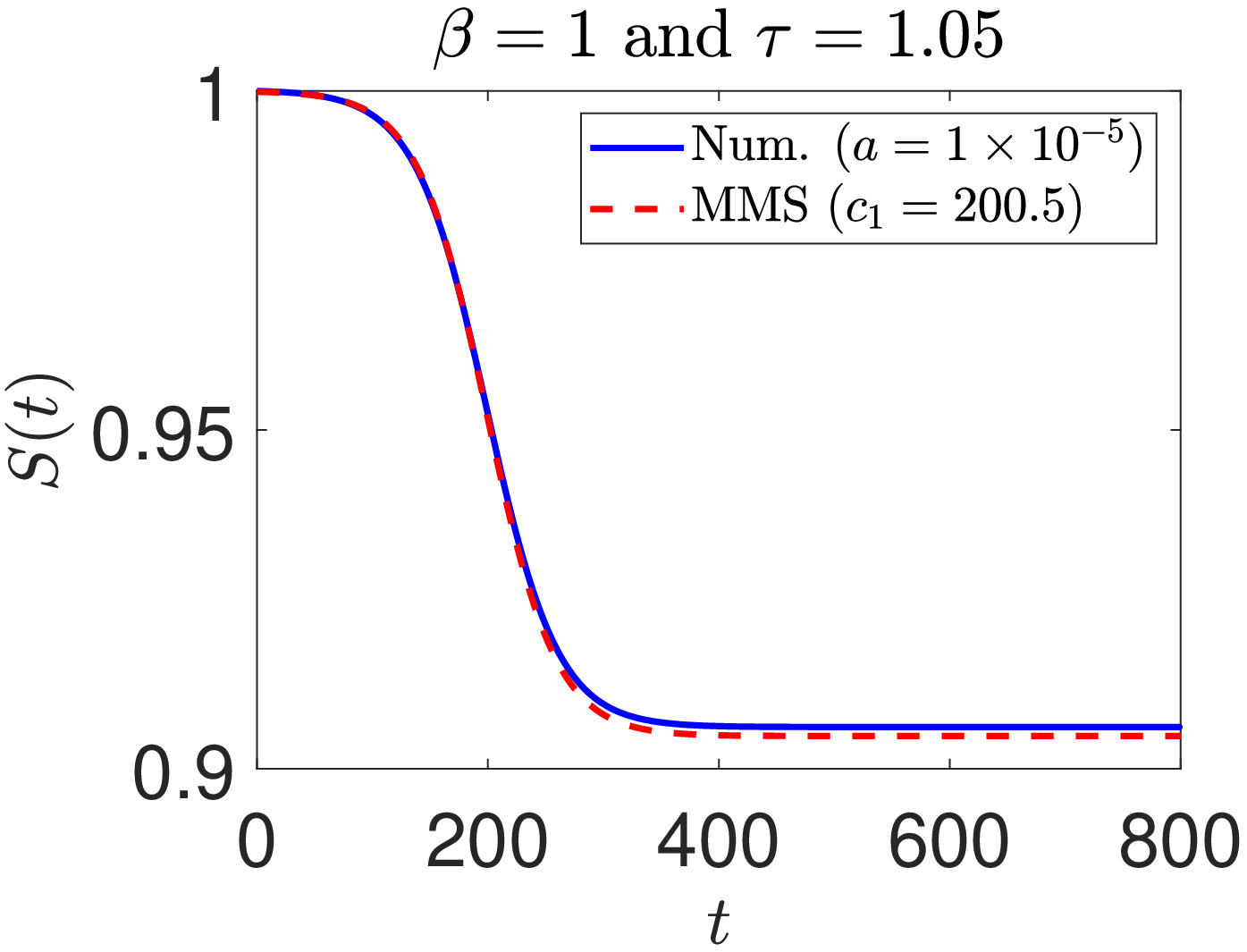}\label{Figure4b}}
	\end{center}
	\caption{Comparison between numerical solution and asymptotic (method of multiple scales, or MMS) solution for (a) $\beta=1$, $\tau=1.025$, $a=1\times10^{-5}$ (numerical) and $c_1=310$ (MMS); (b) $\beta=1$, $\tau=1.05$,  $a=1\times10^{-5}$ (numerical) and $c_1=200.5$ (MMS). The initial condition used for integrating equation~(\ref{Pdotnondim}) is $P(t) = a \left(1+\frac{t}{1+\tau}\right),\, t\le 0$. The arbitrary time-shift $c_1$ of the multiple scales solution (equation~(\ref{finsol2})) is chosen here to obtain a good visual match.}
	\label{Figure4}
\end{figure}

The foregoing results indicate that the $P(t)$ solution starts asymptotically from zero at $t = -\infty$ and grows monotonically if $\beta \tau > 1$, but saturates at a small value when $\beta \tau$ slightly exceeds unity.
We can develop an asymptotic solution for the case where
\begin{equation}
\beta = \frac{1}{\tau} + \epsilon, \quad 0 < \epsilon \ll 1.
\label{epsdef}
\end{equation}
We will use the method of multiple scales~\cite{Kevorkian1995,das2002multiple,nayfeh2008order,hinchperturbation}. We rewrite equation~(\ref{Pdotnondim}) as
\begin{equation}
\dot P(t) =  e^{- \left ( \frac{1}{\tau}+\epsilon \right ) P(t-1-\tau)} - e^{- \left ( \frac{1}{\tau}+\epsilon \right ) P(t-1)}
\label{mms1}
\end{equation}
and note that the $\epsilon = 0$ case is on the stability boundary for $P(t)=0$.
We introduce three time scales for a second order expansion,
\begin{equation}
T_0 = t, \quad T_1 = \epsilon t, \quad T_2 = \epsilon^2 t,
\label{scales}
\end{equation}
think of $P(t)$ as $P(T_0, T_1, T_2)$ with a slight abuse of notation, and observe that the time derivative is to be interpreted as
\begin{equation}
\dot P = \frac{\partial P}{\partial T_0} + \epsilon \frac{\partial P}{\partial T_1} + \epsilon^2 \frac{\partial P}{\partial T_2} + \cdots.
\label{Pmms}
\end{equation}
Further, a delayed quantity such as
$P(t-\Delta)$ is to be interpreted as
\begin{equation}
P(t-\Delta) = P(T_0-\Delta,T_1 - \epsilon \Delta,T_2 - \epsilon^2 \Delta),
\label{Pdelay}
\end{equation}
where due to the smallness of $\epsilon$, Taylor series expansions in $\epsilon$ can be used for the second and third arguments, but not the first argument, i.e.,
\begin{eqnarray}
P(t-\Delta) &= &P(T_0-\Delta,T_1,T_2) - \epsilon \Delta \frac{\partial P(T_0-\Delta,T_1,T_2)}{\partial T_1} +
\frac{\epsilon^2 \Delta^2}{2} \frac{\partial^2 P(T_0-\Delta,T_1,T_2)}{\partial T_1^2}\nonumber \\
&-& \epsilon^2 \Delta \frac{\partial P(T_0-\Delta,T_1,T_2)}{\partial T_2} + 
{\cal O}(\epsilon^3).
\label{TaylorP}
\end{eqnarray}
Finally, $P$ itself is to be expanded as
\begin{equation}
P= \epsilon P_0 + \epsilon^2 P_1 + \epsilon^3 P_2 + \cdots
\label{Pappr}
\end{equation}
where higher order terms in the expansion would require retention of still slower time scales.
This much is routine, and yields an equation of the form (using the symbolic computation software Maple; note that the leading order term is ${\cal O}(\epsilon)$):
\begin{equation}
\left( {\frac {\partial }{\partial T_0}}{ P_0} \left( T_0,T_1,T_2
\right) +{\frac {{P_0} \left( T_0-1-\tau,T_1,T_2 \right) }{\tau}}-{
	\frac {{ P_0} \left( T_0-1,T_1,T_2 \right) }{\tau}} \right) \epsilon  
+
L_2 \epsilon^2 + L_3 \epsilon^3
+ \cdots = 0,
\label{mms2}
\end{equation}
where two long expressions have been written simply as $L_2$ and $L_3$ (details omitted for brevity). At ${\cal O}(\epsilon)$ we have
\begin{equation}
{\frac {\partial }{\partial T_0}}{ P_0} \left( T_0,T_1,T_2
\right) +{\frac {{P_0} \left( T_0-1-\tau,T_1,T_2 \right) }{\tau}}-{
	\frac {{ P_0} \left( T_0-1,T_1,T_2 \right) }{\tau}} = 0,
\label{mmszero}
\end{equation}
for which we adopt the solution (based on previous observations; and also upon rejecting fast-varying exponential decaying terms~\cite{das2002multiple,nayfeh2008order})
\begin{equation}
P_0(T_0,T_1,T_2) = A(T_1,T_2), 
\label{mms3}
\end{equation}
i.e., $P_0$ is a constant on the fast or $T_0$ time scale. Inserting equation~(\ref{mms3}) into equation~(\ref{mms2}), we obtain at ${\cal O}(\epsilon^2)$,
\begin{equation}{\frac {\partial }{\partial T_0}}{ P_1} \left( T_0,T_1,T_2
\right) +{\frac {{P_1} \left( T_0-1-\tau,T_1,T_2 \right) }{\tau}}-{
	\frac {{ P_1} \left( T_0-1,T_1,T_2 \right) }{\tau}} = 0,
\label{mms4}
\end{equation}
with terms containing $A(T_1,T_2)$ canceling each other out at this order. This means $A(T_1,T_2)$
remains indeterminate at this order, and we are free to choose 
\begin{equation}
P_1 \left( T_0,T_1,T_2 \right) = 0,
\end{equation}
because it adds nothing new to the already retained $P_0$. Inserting the above into 
equation~(\ref{mms2}), we obtain at ${\cal O}(\epsilon^3)$,
\begin{equation}
{\frac {\partial }{\partial T_0}}{ P_2} \left( T_0,T_1,T_2
\right) +{\frac {{P_2} \left( T_0-1-\tau,T_1,T_2 \right) }{\tau}}-{
	\frac {{ P_2} \left( T_0-1,T_1,T_2 \right) }{\tau}} =  L_3,
\end{equation}
where $L_3$ is a long expression independent of $T_0$ and containing the function $A(T_1,T_2)$ along with its $T_1$-derivatives only. Now, appealing to the required boundedness of $P_2$ (which corresponds to removal of secular terms, and can also be viewed as a choice that allows the approximation to stay valid for a longer time), we insist that
$L_3 = 0$. Further, since $T_2$ derivatives of $A$ do not appear, we set $A$ back to a function of $T_1$ alone (no contradiction up to this order). In this way, we finally obtain
\begin{equation}
\left ( 1 + \frac{\tau}{2} \right ) A'' - \tau A' + \frac{AA'}{\tau} = 0,
\end{equation}
where we note that $\tau$ is a positive parameter, $A$ is a function of $T_1$, and primes denote $T_1$-derivatives.
The above differential equation is to be solved as a function of $T_1$, with the initial condition $A(-\infty) = 0$. The solution turns out to be
\begin{equation}
A = \tau^2 \left \{ 1 + \tanh \left ( \frac{\tau (T_1 - c_0)}{\tau+2} \right ) \right \},
\end{equation}
where $c_0$ is an indeterminate constant that allows time-shifting (recall figure~\ref{Figure2b} and the related discussion).
Inserting $T_1 = \epsilon t$, and then (recall equation~(\ref{epsdef}))
\begin{equation}
\epsilon = \beta- \frac{1}{\tau},
\end{equation}
we finally obtain the leading order approximation for the entire solution, for $\beta \tau$ slightly greater than unity, as
\begin{equation}
P(t) \sim (\beta \tau -1) \tau \left \{ 1 + \tanh \left ( \frac{(\beta \tau -1)(t-c_1)}{\tau+2} \right ) \right \},
\label{finsol}
\end{equation}
where $c_1$ is an undetermined constant.
The above solution saturates, as $t \rightarrow \infty$, at
\begin{equation}
P(\infty) \sim 2\tau(\beta \tau -1),
\label{Pinf}
\end{equation}
which matches equation~(\ref{Cap}) upon noting that we can replace $\frac{C}{\beta}$ with $C \tau$ with no errors introduced at leading order.
At the same order of approximation, we can also write
\begin{equation}
P(t) \sim  \frac{\beta \tau -1}{\beta} \left \{ 1 + \tanh \left ( \frac{\beta (\beta \tau -1)(t-c_1)}{1+2 \beta} \right ) \right \},
\label{finsol1}
\end{equation}
whence
\begin{equation}
S(t) \sim e^{\displaystyle -(\beta \tau -1) \left \{ 1 + \tanh \left ( \frac{\beta (\beta \tau -1)(t-c_1)}{1+2 \beta} \right ) \right \}},
\label{finsol2}
\end{equation}
where we have used the `$\sim$' notation because this is an asymptotic approximation.
A numerical example is given in figure~\ref{Figure4} for $\beta\tau =1.025$ ($\beta=1$ and $\tau=1.025$) and $\beta\tau =1.05$ ($\beta=1$ and $\tau=1.05$). The match is good, but deteriorates for larger values of $\beta \tau$.

It may be noted that, for such weak growth where only a small fraction of the total population gets infected before the pandemic runs its course, by equation~(\ref{finsol1}),
\begin{equation}
\label{Sinf2}
S(\infty) \sim 1 - 2(\beta \tau -1).
\end{equation}
In comparison (recall equation~(\ref{Smax})), $S$ could in principle be stable at a value as high as
\begin{equation}
\label{Sstar}
S^* = \frac{1}{\beta \tau} = \frac{1}{1+(\beta \tau - 1)} \sim 1 - (\beta \tau -1),
\end{equation}
i.e., the uncontrolled and constant-$\beta$ solution infects about twice as many people as seems strictly necessary; we will return to this in section \ref{tvb}.

We will next develop a long-wave approximation that performs better for somewhat larger $\beta \tau$.

\subsection{Long-Wave Approximation for Moderate Growth}
The advantage of using an asymptotic method like the method of multiple scales is that we have formal validity as $\epsilon \rightarrow 0$. We are reassured that the solution we are seeking does in fact exist, and has approximately the shape obtained as the leading order approximation. However, in the present case, for somewhat larger $\epsilon$ the solution is not very accurate; moreover, proceeding to higher orders leads to long expressions that seem difficult to simplify usefully. Therefore, encouraged by our asymptotic solution, we now develop a more informal but more accurate approximation. In particular, we try a long-wave (LW) approximation as follows.

We suppose that there is a ``long'' scale (technically a time scale, for this problem) which we shall call $L$, such that
\begin{equation}
P(t) = \hat P \left ( \frac{t}{L} \right ), \quad L \gg 1.
\label{anstz}
\end{equation}
Let 
\begin{equation}
\xi = \frac{t}{L}.
\end{equation}
Now equation~(\ref{Pdotnondim}) becomes
\begin{equation}\frac{{\hat P}'(\xi)}{L} = e^{-\beta \hat P \left ( \xi - \frac{1+\tau}{L} \right )} - e^{-\beta \hat P \left ( \xi - \frac{1}{L} \right )}.
\end{equation}
Expanding the above in a series for large $L$, retaining terms up to ${\cal O}(L^{-2})$, and solving for
$\hat P''$, we obtain
\begin{equation}
\hat P'' = \beta \hat P'^2 + \frac{2L \hat P'}{\tau+2} \left ( 1 - \frac{e^{ \beta \hat P} }{\beta \tau} \right ).
\label{e1} 
\end{equation}
We note that on the right hand side the second term is dominant because it contains the large parameter $L$, while the first term does not. The left hand side has the highest derivative and cannot be dropped without changing the order of the differential equation. For these reasons, a further approximation to equation~(\ref{e1}) is
\begin{equation}
\label{e2} \hat P'' =  \frac{2L \hat P'}{\tau+2} \left ( 1 - \frac{e^{ \beta \hat P} }{\beta \tau} \right ).
\end{equation}
Equation~(\ref{e2}) is integrable, and yields
\begin{equation}
\hat P' = \frac{2L}{\tau+2} \left ( \hat P - \frac{e^{ \beta \hat P} }{\beta^2 \tau} \right ) + C_0,
\end{equation}
where $C_0$ is an integration constant. The initial condition of interest is $\hat P' = 0$ and $\hat P = 0$ as
$\xi \rightarrow - \infty$, whence we obtain
\begin{equation}
\label{e3} \hat P' = \frac{2L}{\tau+2} \left ( \hat P - \frac{e^{ \beta \hat P} }{\beta^2 \tau} \right ) + \frac{2L}{\beta^2 \tau(\tau+2)}.
\end{equation}

We now return to equation~(\ref{e1}), where we can insert equation~(\ref{e3}) into the non-dominant term as an approximation. A simple way to do it is to replace the first term on the right hand side with an approximation that is integrable, i.e.,
\begin{equation} 
\beta \hat P'^2 = \beta \hat P' \left \{ \frac{2L}{\tau+2} \left ( \hat P - \frac{e^{ \beta \hat P} }{\beta^2 \tau} \right ) + \frac{2L}{\beta^2 \tau(\tau+2)} \right \}.
\end{equation}
This gives
\begin{equation} 
\hat P'' = \beta \hat P' \left \{ \frac{2L}{\tau+2} \left ( \hat P - \frac{e^{ \beta \hat P} }{\beta^2 \tau} \right ) + \frac{2L}{\beta^2 \tau(\tau+2)} \right \} + \frac{2L \hat P'}{\tau+2} \left ( 1 - \frac{e^{ \beta \hat P} }{\beta \tau} \right ).
\end{equation}
The above is integrable, and after enforcing the initial condition $\hat P' = 0$ and $\hat P = 0$ as
$\xi \rightarrow - \infty$, setting $L=1$ (it was a bookkeeping parameter all along, helping us keep track of which effects are big and which are small), and dropping the hat, we obtain the informal long-wave approximation
\begin{equation}
\frac{dP}{dt} = \frac{(2+\beta P)P}{\tau+2} + \frac{2P}{\beta \tau (\tau+2)} + \frac{4}{\beta^2 \tau ( \tau+2)} \left ( 1 - e^{\beta P} \right ).
\label{e4}
\end{equation}

\begin{figure}[htpb!]
	\begin{center}
		\subfigure[]{\includegraphics[width=0.48\textwidth]{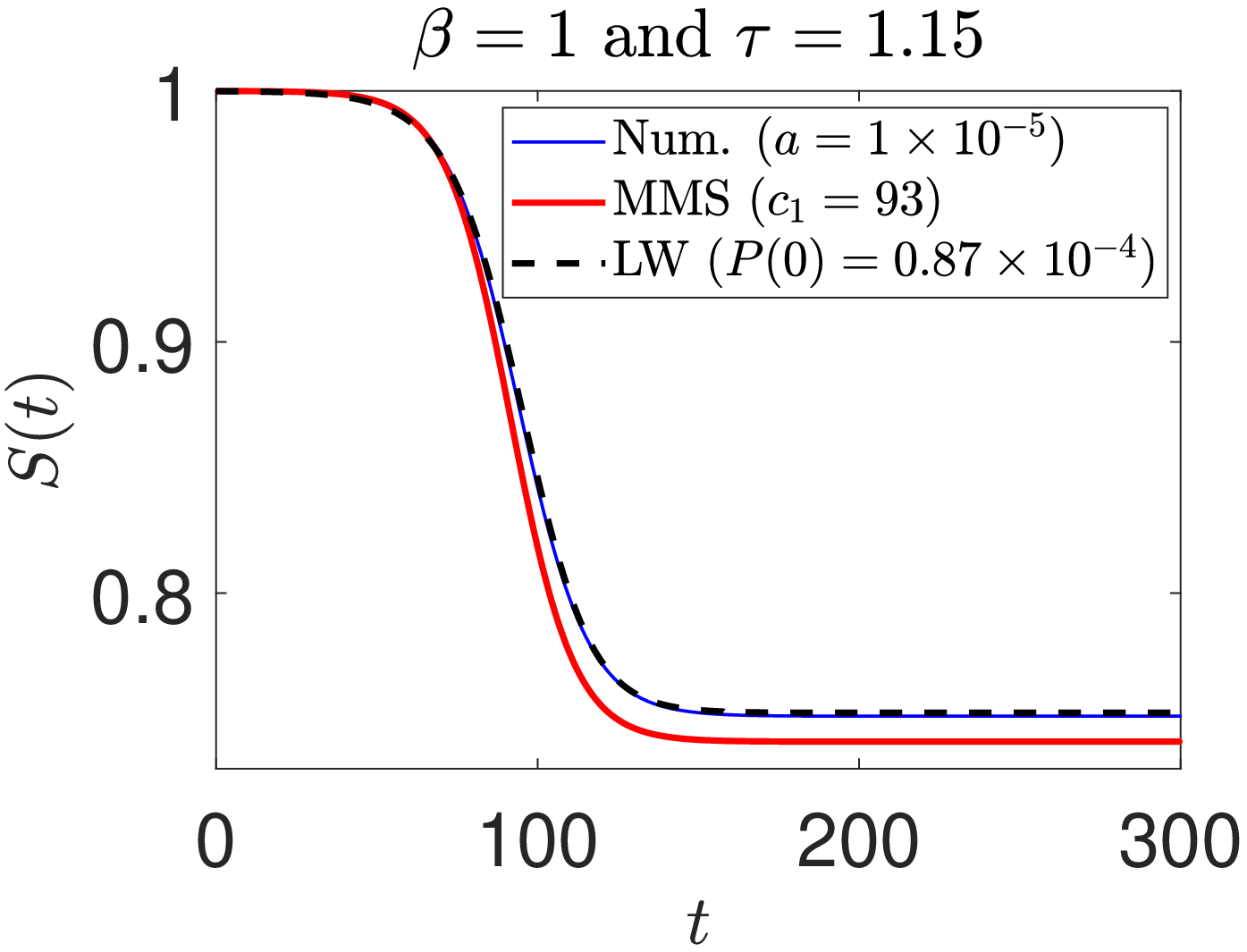}\label{Figure5a}}
		\subfigure[]{\includegraphics[width=0.48\textwidth]{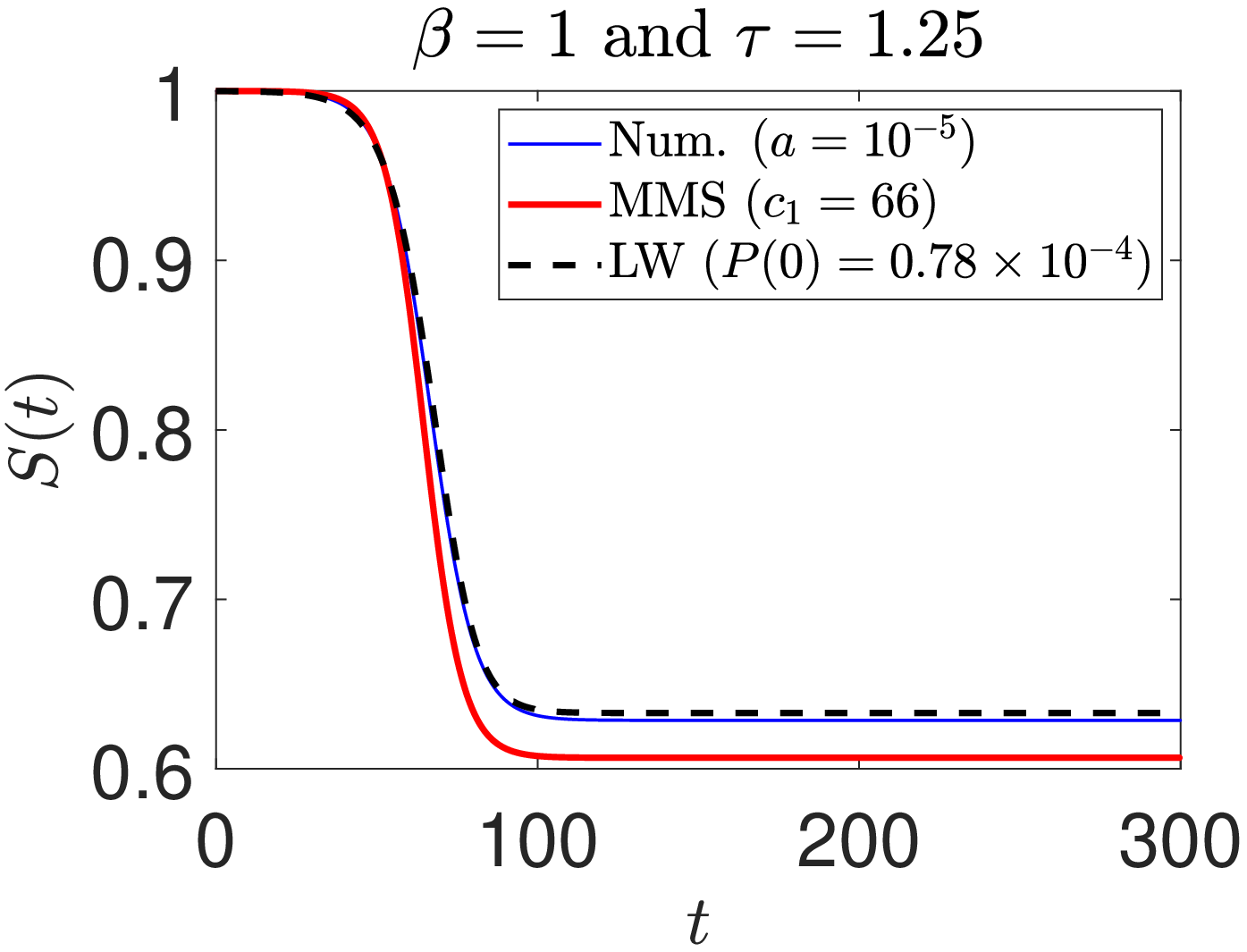}\label{Figure5b}}
	\end{center}
	\caption{Comparison between numerical, multiple scales (MMS), and long-wave solutions for (a) $\beta=1$, $\tau=1.15$, $a=1\times10^{-5}$ (numerical),  $c_1=93$ (MMS), and $P(0)=0.87\times 10^{-4}$ (long-wave);  (b) $\beta=1$, $\tau=1.25$, $a=1\times10^{-5}$ (numerical),  $c_1=66$ (MMS), and $P(0)=0.78\times 10^{-4}$ (long-wave). The initial condition used for integrating equation~(\ref{Pdotnondim}) is $P(t) = a \left(1+\frac{t}{1+\tau}\right),\, t\le 0$. The initial condition $P(0)$ is used to integrate the long-wave equation~(\ref{e4}). The coefficient $c_1$ is used in the multiple scales solution (see equation~(\ref{finsol2})).}
	\label{Figure5}
\end{figure}

We mention that equation~(\ref{e4}) is indeed a significant simplification, because it replaces a DDE (infinite-dimensional phase space) with a first order ODE (one-dimensional phase space). It also applies to a specific solution, all the way from infinitesimal initiation to final saturation. The freedom in a single initial condition that it offers is equivalent to simply the arbitrary
time-shift already noted in the multiple scales solution. While it cannot be solved explicitly in closed form, its solution can be formally expressed in implicit form using an indefinite integral (omitted for brevity).

The qualitative behavior of the approximation in equation~(\ref{e4}) may be checked easily for small positive $P$ by linearizing the right hand side to obtain
\begin{equation}
\dot P = \frac{2(\beta \tau - 1)}{\beta \tau (\tau + 2)}P,
\end{equation} 
which is consistent with the earlier result that growth requires $\beta \tau > 1$ (inequality~(\ref{sat0})).
In figure~\ref{Figure5}, numerical solutions for $\beta \tau$ somewhat greater than unity are shown. We consider
$\beta \tau = 1.15$ (see figure~\ref{Figure5a}) and $\beta \tau = 1.25$ (see figure~\ref{Figure5b}). The numerically obtained long-wave solutions match well with full numerical solutions of the original DDE; in fact, they match significantly better than the multiple scales solution. 

Finally, a direct analytical comparison with the multiple-scales solution can be made by expanding the right hand side of equation~(\ref{e4}) in a power series for small $P$, retaining upto quadratic terms. That equation can be solved in closed form, and gives a hyperbolic tangent solution as well, which initially looks a little different from the multiple scales solution:
\begin{equation}
\label{LW_tanh}
P(t) = \frac{\beta\tau - 1}{\beta(2 - \beta \tau)} \left \{ 1 + \tanh \left ( \frac{(\beta \tau - 1)(t - c_1)}{\beta \tau(\tau+2)} \right ) \right \}.
\end{equation}
However, noting that the multiple scales solution was for $\beta \tau$ close to unity, if we replace $(2 -\beta \tau)$ with $2-1=1$, and replace
$\beta \tau (\tau + 2)$ with $1 \cdot \left ( \frac{1}{\beta} + 2 \right )$, then we recover equation~(\ref{finsol1}). This is not surprising because for
$\beta \tau$ slightly greater than unity, the long-wave approximation is asymptotic as well. The approximation is only informal (as opposed to asymptotic) when we use it for arbitrary values of $\beta$ and $\tau$, with $\beta \tau$ not close to unity.

This concludes our study of the $p=1$ and $\lambda = 0$ subcase, which is interesting both because it is a reasonable limit and because it permits any constant $P$ as an equilibrium. The latter is not true for general parameter values, to which we turn next. Encouraged by the simplicity of the long-wave approximation as opposed to the multiple scales solution, we try only the former.

\section{Long-Wave Approximation for General Parameter Values}

We now take up equation~(\ref{Pdotnon}), reproduced below:
$$
\dot P(t) =  \bar{p}e^{-\beta P(t-1-\tau)}-e^{-\beta P(t-1)}-\gamma P(t)+1-\bar{p}.
$$
Proceeding with the same ansatz as equation~(\ref{anstz}), expanding up to second order, and setting $L$ to unity, we obtain
\begin{figure}[htpb!]
	\begin{center}
		\subfigure[]{\includegraphics[width=0.48\textwidth]{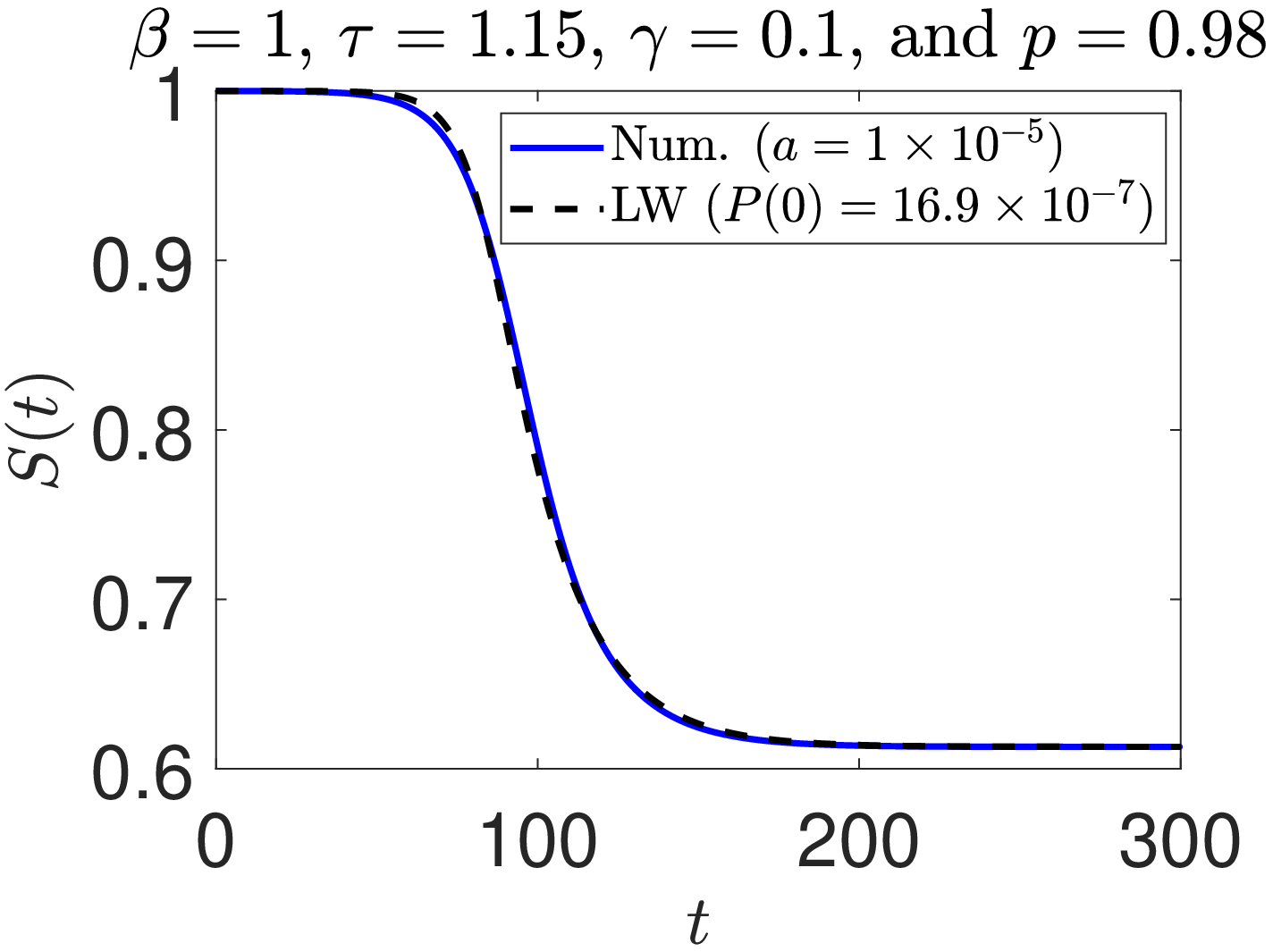}\label{Figure6a}}
		\subfigure[]{\includegraphics[width=0.48\textwidth]{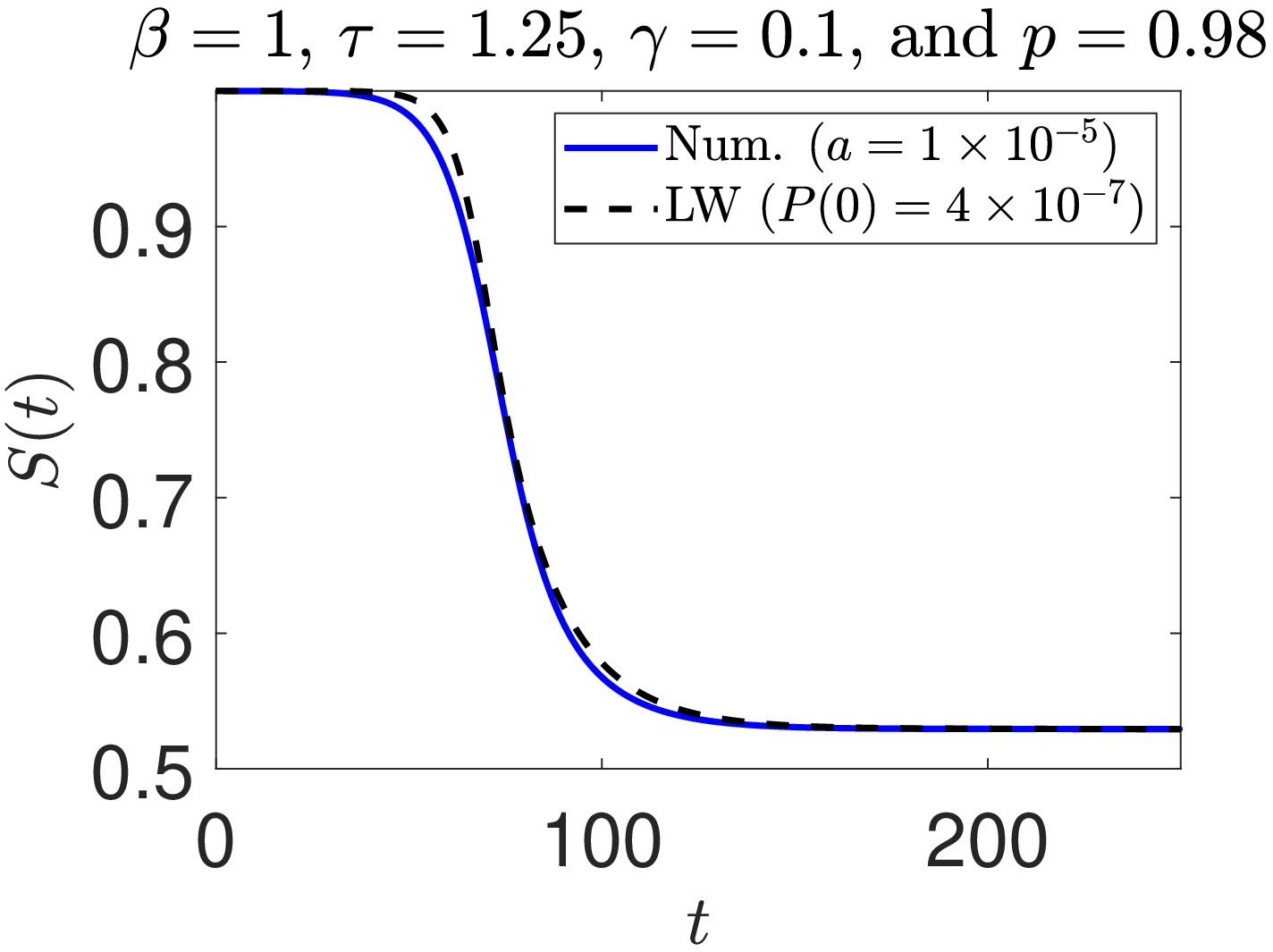}\label{Figure6b}}
	\end{center}
	\caption{Comparison between numerical and long-wave solutions for (a) $\beta=1$, $\tau=1.15$, $a=1\times10^{-5}$ (numerical),  and $P(0)=16.9\times 10^{-7}$ (long-wave);  (b) $\beta=1$, $\tau=1.25$, $a=1\times10^{-5}$ (numerical),   and $P(0)=4\times 10^{-7}$ (long-wave). The initial condition used for integrating equation~(\ref{Pdotnondim}) is $P(t) = a \left(1+\frac{t}{1+\tau}\right),\, t\le 0$. The initial condition $P(0)$ reported in the figure and $\dot{P}(0)=0$ are used to integrate the long-wave equation~(\ref{LWgen}).}
	\label{Figure6}
\end{figure}
\begin{equation}
\label{LWgen}
\ddot P = \beta \dot P^2 + \frac{2}{\mu} \left ( \bar p (1 + \tau) -1 - \frac{e^{\beta P}}{\beta} \right ) \dot P 
+ \frac{2}{\beta \mu} \left  ( \bar p -1 +  ( 1 - \bar p - \gamma P  ) e^{\beta P} \right ),
\end{equation}
where
\begin{equation}
\label{eqmu} \mu = \bar p (1+\tau)^2 -1.
\end{equation}
A few things may be noted here. First, assuming that $\tau$ and $\bar p$ are not too small, we assume $\mu > 0$ in equation~(\ref{eqmu}). Second, we were able to use a trick for the $p=1$ and $\gamma = 0$ case to reduce the order of the approximating long-wave differential equation; for those parameter values, the last term on the right hand side in equation~(\ref{LWgen}) becomes zero. Here, we have had to retain the second order ordinary differential equation. Third, if there is a nonzero positive
$P$ for which equation~(\ref{LWgen}) has an equilibrium solution, then that $P$ must satisfy
\begin{equation}
\bar p -1 +  ( 1 - \bar p - \gamma P  ) e^{\beta P} = 0,
\label{longss}
\end{equation}
which is equivalent to equation~(\ref{equilval}). This means the steady state $P$ in this general case will be exactly correct, unlike the previous subcase
of $p=1$ and $\gamma = 0$. However, note that this unique limiting $P$ is for the specific solution that starts asymptotically at zero, at $t = -\infty$. Since we now have a second order differential equation in the long-wave approximation, trying to approximate what is purportedly a single solution, we have a minor dilemma in interpreting the two degrees of freedom
available in initial conditions for equation~(\ref{LWgen}). One of those corresponds to an arbitrary time-shift, as noted earlier. That leaves one other initial condition. Since our derivation has not identified what this initial condition should be, we expect that it should be inconsequential. This does turn out to be the case, as seen next.

For small $P$ and small $\dot P$, equation~(\ref{LWgen}) can be linearized to give
\begin{equation}
\label{LWgenlin}
\ddot P = \frac{2}{\mu} \left ( \bar p (1 + \tau) -1 - \frac{1}{\beta} \right ) \dot P + \frac{2}{\beta \mu} ( \beta ( 1 - \bar p) - \gamma) P.
\end{equation}
For stability, coefficients of both $\dot P$ and $P$ above need to be negative. If the coefficient of $\dot P$ is positive while the coefficient of
$P$ remains negative, then growing oscillations are predicted; such solutions are non-physical for our application, because $P$ must be monotonic. For monotonic growth, it is the coefficient of $P$ that must be positive. This leads to the condition for existence of such solutions,
\begin{equation} \label{stb1}
\frac{\beta}{\gamma} ( 1 - \bar p) > 1,
\end{equation}
which matches the (loss of) stability condition of Young {\em et al.}~\cite{young2019consequences} as well as our equation~(\ref{reduce}). In other words, the condition for a  solution growing from zero is exactly the same condition for the existence of another equilibrium for strictly positive $P$; and this condition, though obtained here from the long-wave approximation, matches the theoretical result exactly because it is near this stability boundary that the long-wave approximation is asymptotic. Finally, when the coefficient of $P$ in equation~(\ref{LWgenlin}) is indeed positive, then the two characteristic roots are real: one is positive and one is negative. The positive root leads to the growing solution, as above. The negative root absorbs the apparently free initial condition, contributes an exponentially decaying term that dies soon, and has no influence on the growing solution provided the initial conditions
are sufficiently early in the outbreak.

A numerical example is shown in figure~\ref{Figure6} for two cases: $\beta\tau=1.15$ and $\beta\tau=1.25$. The parameter $p=0.98$ implies that the probability of detecting infected individuals is not perfect. Further, $\gamma=0.1$ models a small fraction of the infectious population recovering without being quarantined. In the figure, the long-wave solution perfectly matches the final saturation state obtained from numerical integration of the DDE (equation~(\ref{Pdotnondim})), due to the equivalence of equations~(\ref{longss}) and (\ref{equilval}). 

We have so far concentrated on approximations for a specific solution: one that starts from infinitesimal infection levels, changes monotonically but slowly over a long time and accelerates over a relatively short time, to finally saturate at a finite value as the time goes to infinity.

We now consider more general dynamics, starting from less restricted initial conditions, and allowing for a time-varying infection
rate $\beta$.

\section{Time-Varying $\beta$ and Related Policy Implications}
\label{tvb}
As public health policy based on observed spread of the disease, a government may prescribe temporarily greater social distancing, effectively lowering $\beta$. We can then use equations~(\ref{Sdot2a}) and~(\ref{Idot2a}), rewritten here for readability (incorporating
$\bar p$ from equation (\ref{pbareq})):
\begin{eqnarray}
\label{ge10} 
\dot{S}(t)&=&\beta(t) S(t)I(t),\\
\label{ge11}
\dot{I}(t)&=&\beta(t-1) S(t-1)I(t-1) - \bar p \beta(t-1-\tau)  S(t-1-\tau)I(t-1-\tau) -\gamma I(t).
\end{eqnarray}

We first present a six-state Galerkin approximation for equations (\ref{ge10}) and (\ref{ge11}) using Legendre polynomials as basis functions. The ODEs from the Galerkin approximation are shown below (see the appendix for a brief derivation and refer to~\cite{wahi2005Galer,sadath2015galerkin} for details). Writing
$\nu=1+\tau$, $\alpha_{2}=1-\frac{2}{\nu}$ and $\alpha_{3}=\frac{3}{2}\left(1-\frac{2}{\nu}\right)^{2}-\frac{1}{2}$, the reduced order model is 
\begin{eqnarray}
\label{eta1dot}
\dot{\eta}_{1}&=&\frac{2}{\nu}\eta_{2},\\
\label{eta2dot}
\dot{\eta}_{2}&=&\frac{6}{\nu}\eta_{3},\\
\label{eta3dot}
\dot{\eta}_{3}&=&-\frac{2}{\nu}\eta_{2}-\frac{6}{\nu}\eta_{3}-\beta(t)\left(\eta_{1}+\eta_{2}+\eta_{3}\right)\left(\eta_{4}+\eta_{5}+\eta_{6}\right),\\
\label{eta4dot}
\dot{\eta}_{4}&=&\frac{2}{\nu}\eta_{5},\\
\label{eta5dot}
\dot{\eta}_{5}&=&\frac{6}{\nu}\eta_{6},\\
\label{eta6dot}
\dot{\eta}_{6}&=&-\frac{2}{\nu}\eta_{5}-\frac{6}{\nu}\eta_{6}+\beta(t-1)\left(\eta_{1}+\alpha_{2}\eta_{2}+\alpha_{3}\eta_{3}\right)\left(\eta_{4}+\alpha_{2}\eta_{5}+\alpha_{3}\eta_{6}\right)\nonumber\\
&-&\bar{p}\beta(t-\nu)\left(\eta_{1}-\eta_{2}+\eta_{3}\right)\left(\eta_{4}-\eta_{5}+\eta_{6}\right)-\gamma\left(\eta_{4}+\eta_{5}+\eta_{6}\right).
\end{eqnarray}
In the above approximation 
$S(t)=\eta_1(t)+\eta_2(t)+\eta_3(t)$ and $I(t)=\eta_4(t)+\eta_5(t)+\eta_6(t)$. Here, $\beta(t)$ has delays but the state variables do not. Note that we approximate the dynamical system here, and not a specific solution
as we did with the long wave approximation.

\begin{figure}[htpb!]
	\begin{center}
		\subfigure[]{\includegraphics[width=0.48\textwidth]{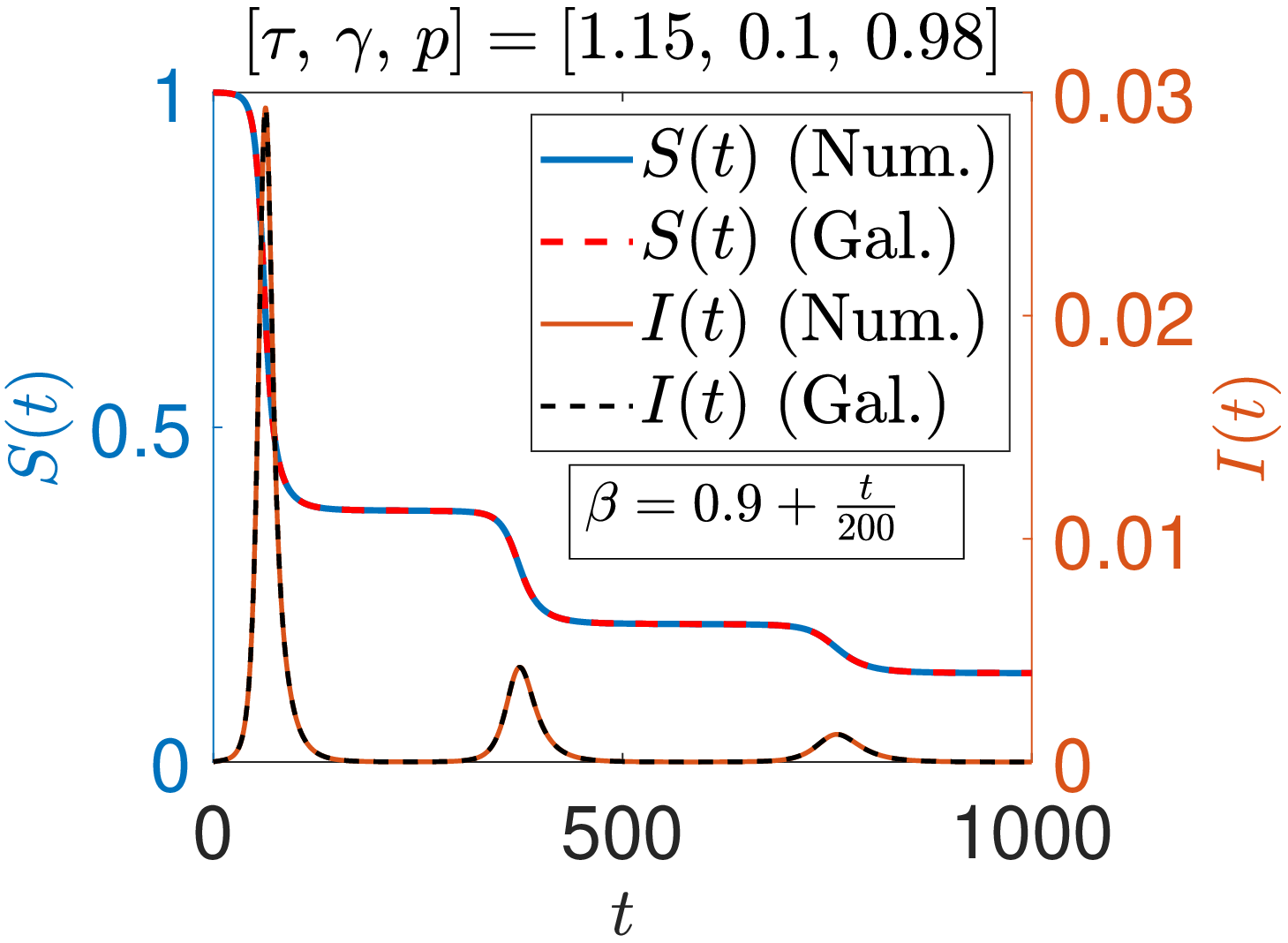}\label{Figure7a}}
		\subfigure[]{\includegraphics[width=0.48\textwidth]{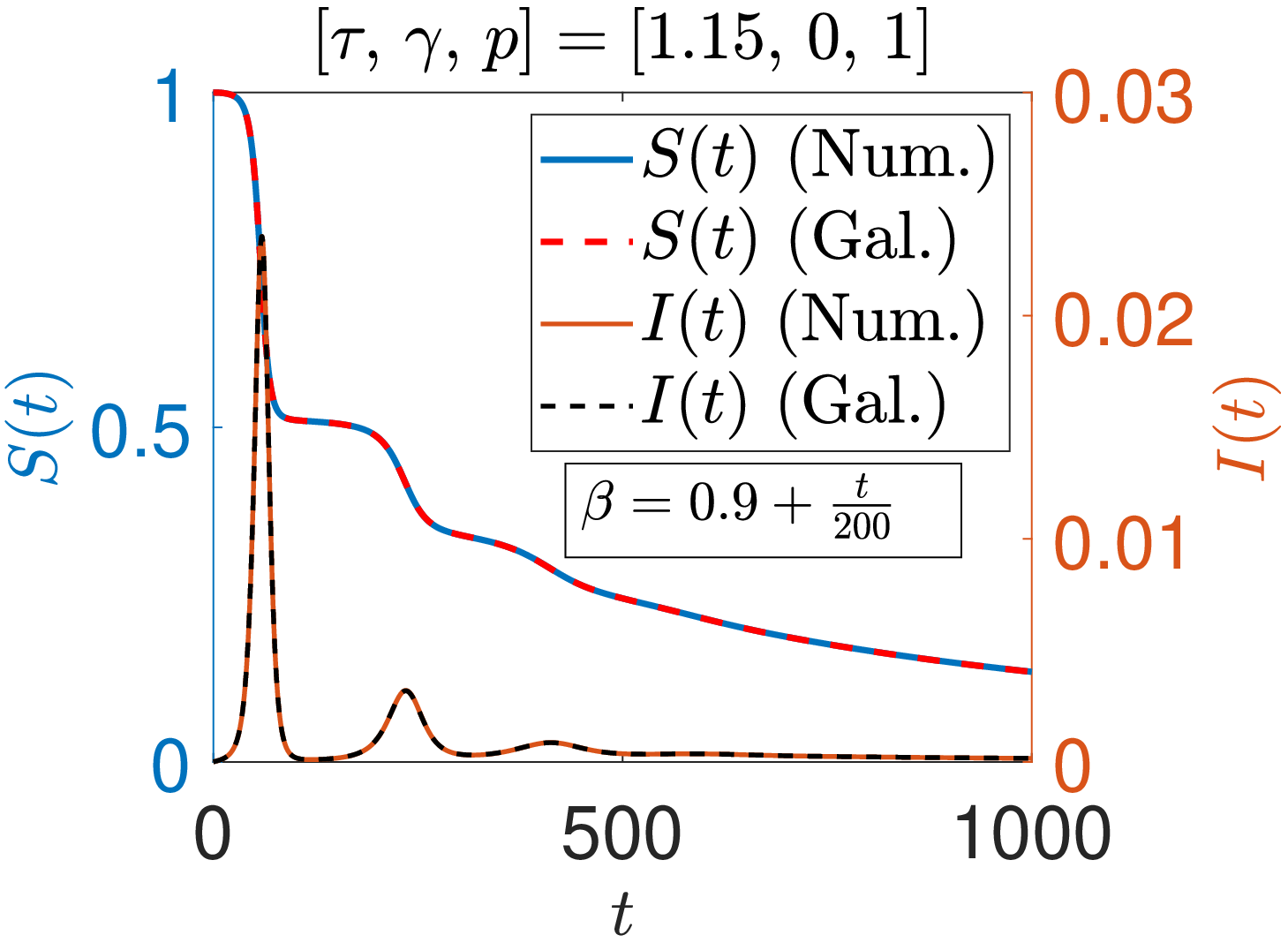}\label{Figure7b}}
		\smallskip
		\subfigure[]{\includegraphics[width=0.48\textwidth]{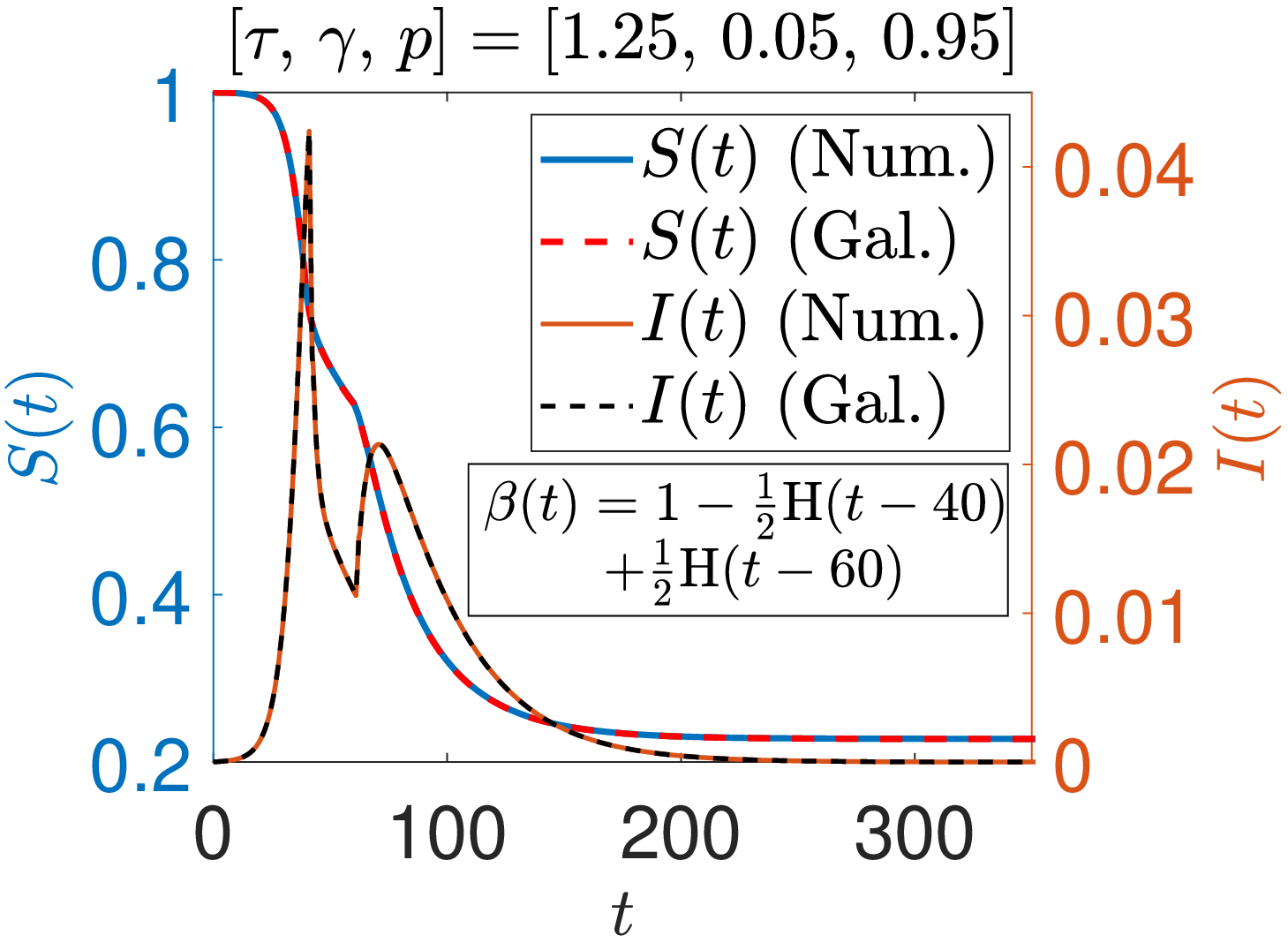}\label{Figure7c}}
		\subfigure[]{\includegraphics[width=0.48\textwidth]{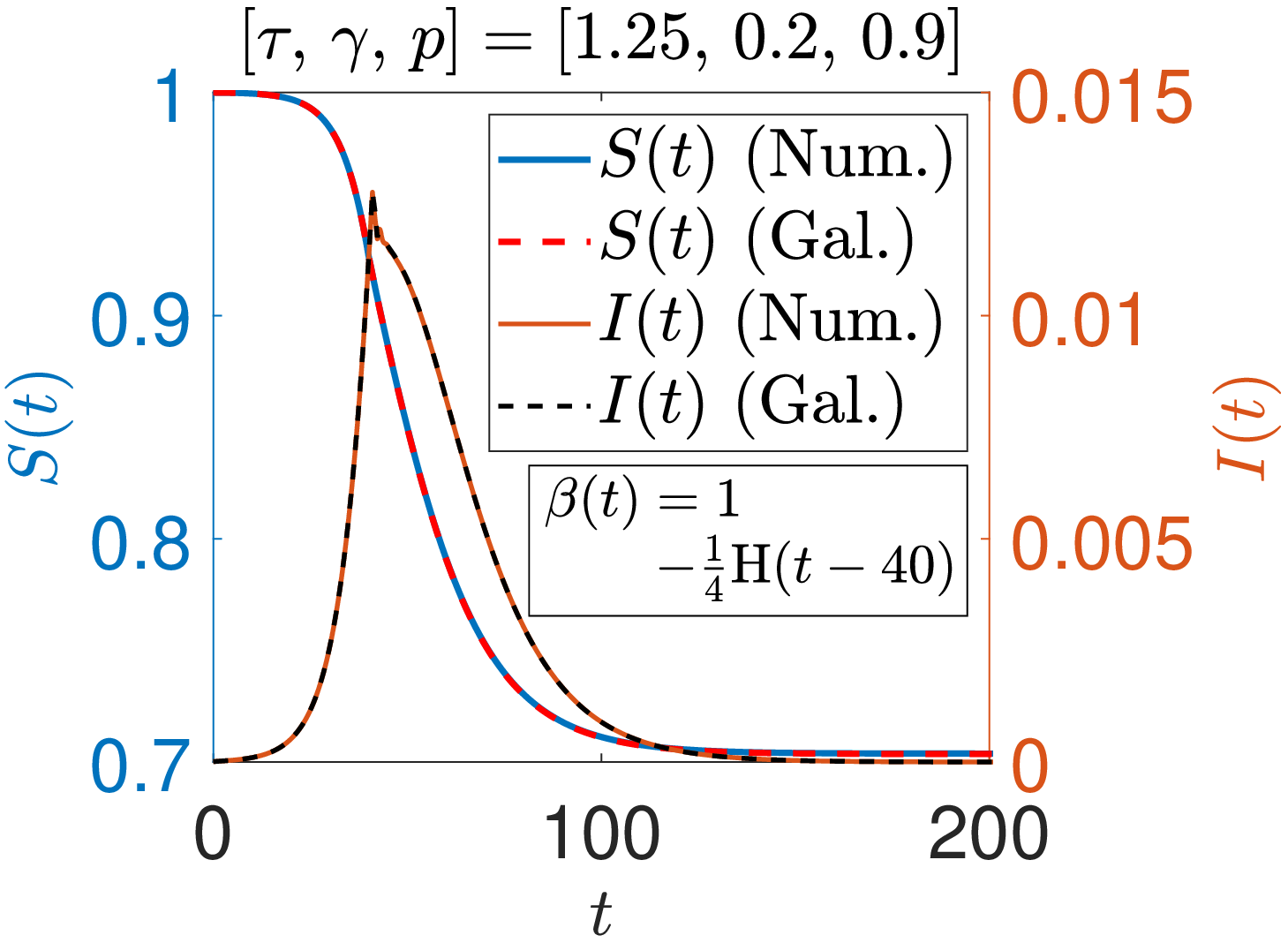}\label{Figure7d}}
		\smallskip
		\subfigure[]{\includegraphics[width=0.48\textwidth]{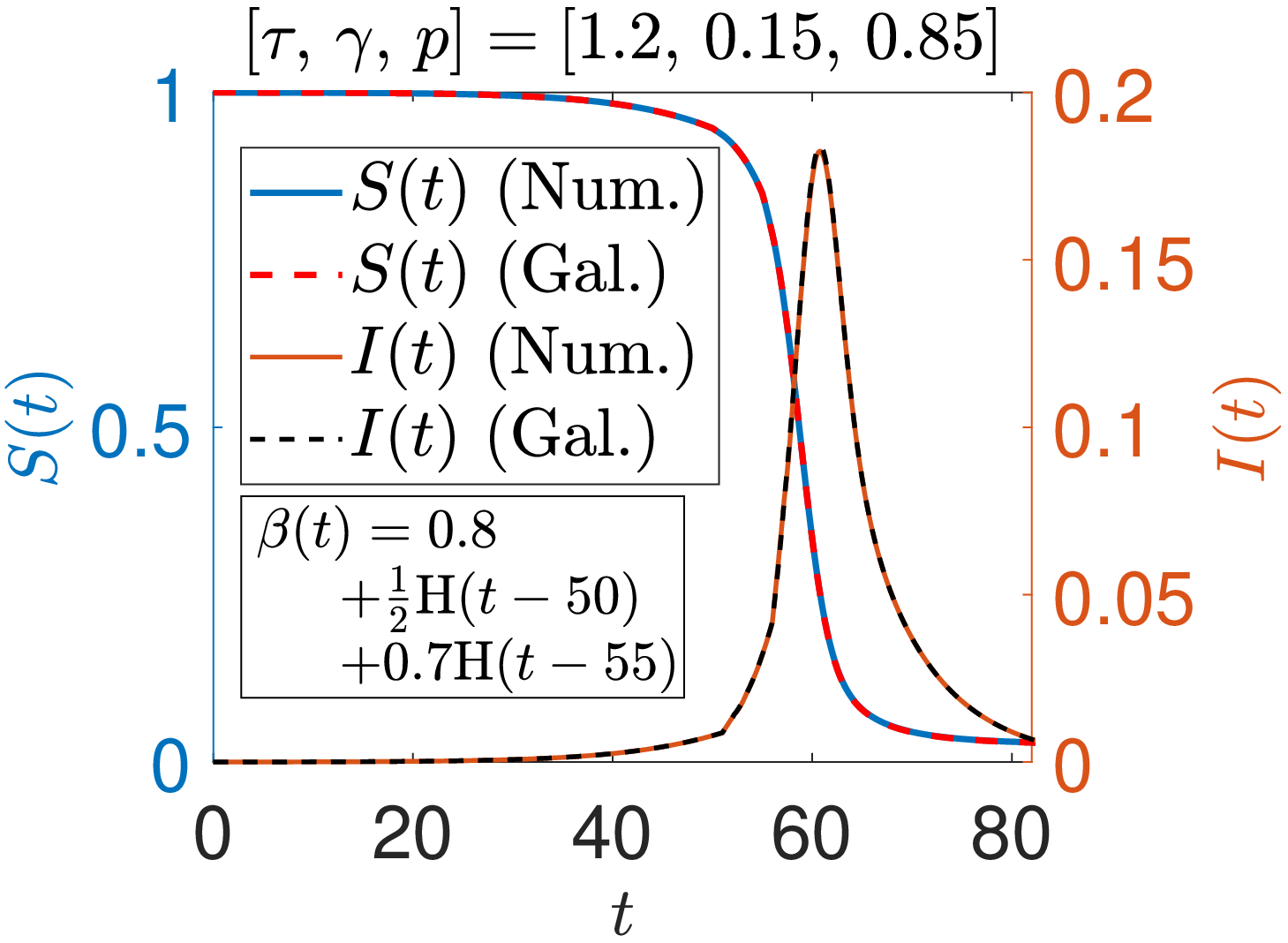}\label{Figure7e}}
		\subfigure[]{\includegraphics[width=0.48\textwidth]{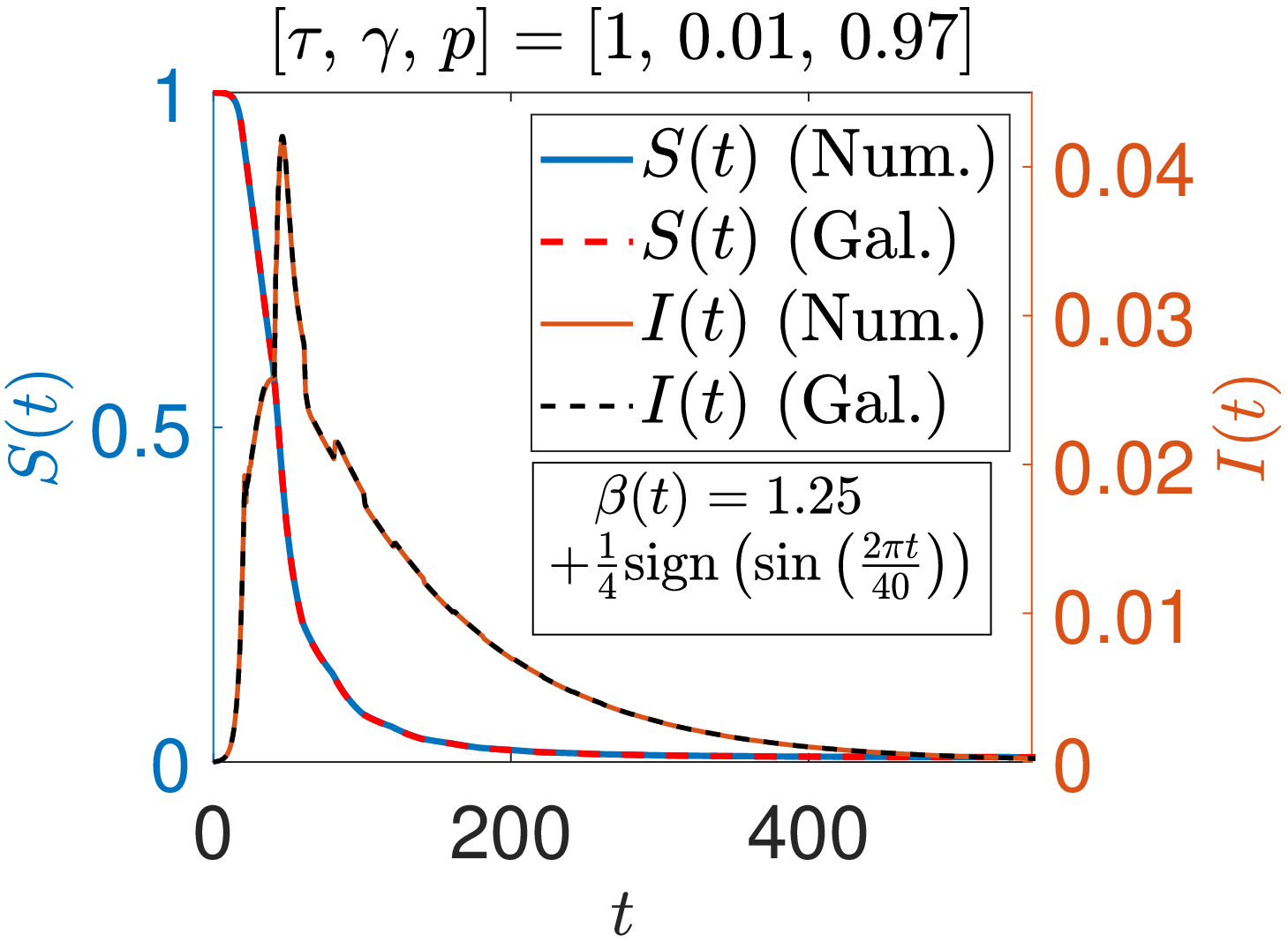}\label{Figure7f}}
	\end{center}
	\caption{Comparison between numerical and Galerkin solutions for various parameters.  $H(t-c) = 1$ 
		if $t>c$, and is zero otherwise. Initial functions used for equations~(\ref{ge10}) and (\ref{ge11}) are $S(t)=1-10^{-5}\left(1+\frac{t}{1+\tau}\right),\,t\le0$ and $I(t)=10^{-5}\left(1+\frac{t}{1+\tau}\right),\,t\le0$. Initial conditions for equations (\ref{eta1dot})-(\ref{eta6dot}) are fitted by Galerkin projection (see the appendix).}
	\label{Figure7}
\end{figure}

The accuracy of the reduced order model is shown in figure~\ref{Figure7}. For each of several sets of parameters, we find an excellent match between numerical solutions of the DDEs and the Galerkin based ODEs. Both continuous and discontinuous $\beta$'s are considered. The reduced order model shows that our
DDEs (equations~(\ref{ge10}) and (\ref{ge11})), though formally infinite-dimensional systems, are {\em effectively} finite-dimensional. The remaining dynamics consists of rapidly decaying components that are soon inconsequential.

Having demonstrated that the dynamics is effectively low-dimensional, we can have greater faith in simple numerical studies that suggest policy implications.
We now turn to such a policy implication. The issue was anticipated in section \ref{restc}, and we now discuss it in detail.

\begin{figure}[htpb!]
	\begin{center}
		\subfigure[]{\includegraphics[width=0.45\textwidth]{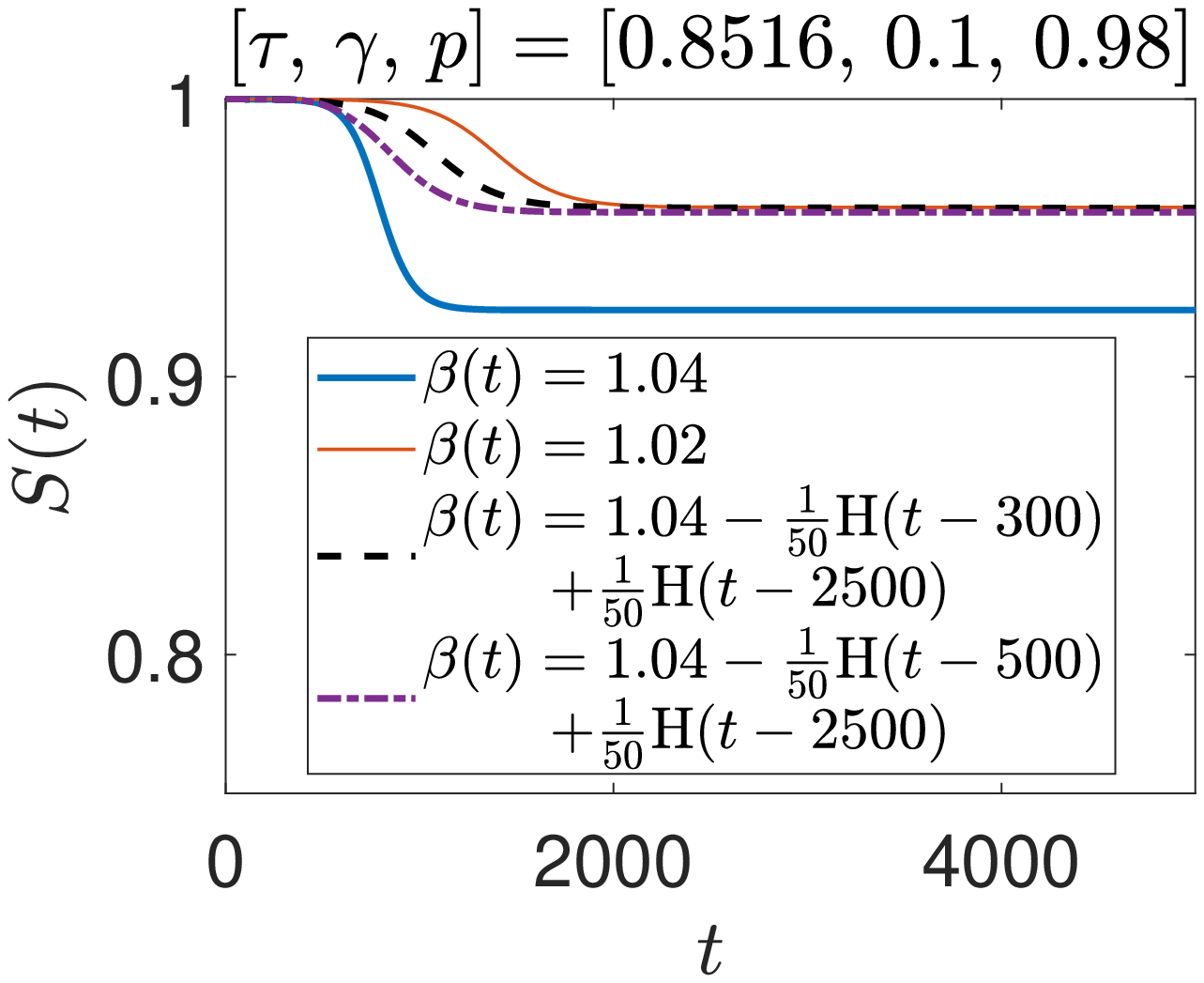}\label{Figure8a}}
		\subfigure[]{\includegraphics[width=0.45\textwidth]{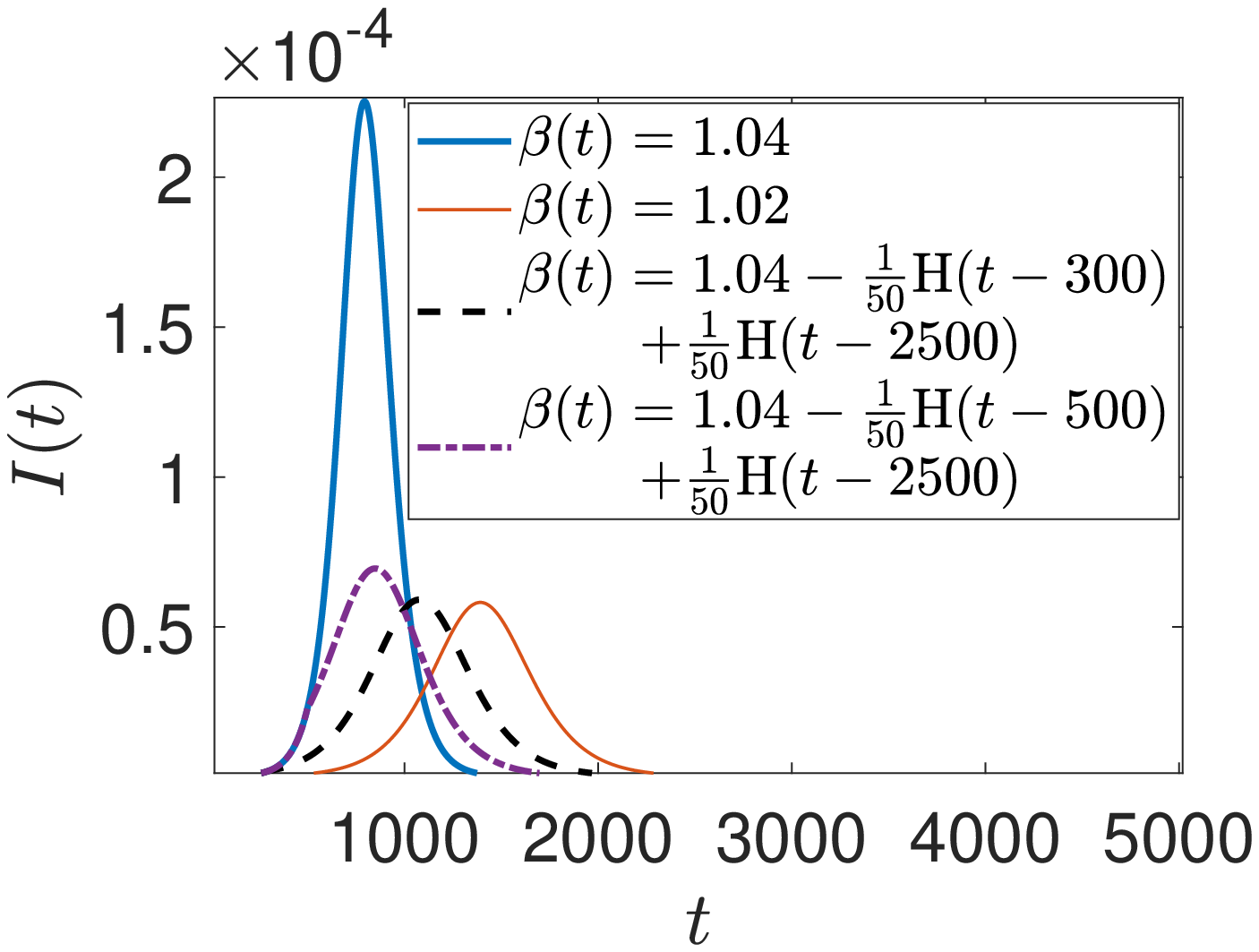}\label{Figure8b}}
	\end{center}
	\caption{Role of time-varying $\beta$. For $\beta(t)=1.04$ held fixed, $7.6\%$ of the population gets infected. With $\beta(t)=1.04-0.02\textrm{H}(t-300)+0.02\textrm{H}(t-2500)$,  $3.9\%$ of the population gets infected and the outbreak proceeds no further. This is practically the same as having $\beta = 1.02$ for all time. When the switch occurs later, with $\beta(t)=1.04-0.02\textrm{H}(t-500)+0.02\textrm{H}(t-2500)$, a slightly larger fraction of the population gets infected. Initial conditions used for equations~(\ref{ge10}) and (\ref{ge11}) were $S(t)=1-1\times10^{-8}\left(1+\frac{t}{1+\tau}\right),\,t\le0$ and $I(t)=1\times10^{-8}\left(1+\frac{t}{1+\tau}\right)$.}
	\label{Figure8}
\end{figure}

We suppose that normal living entails some specific $\beta$, and the public cannot indefinitely maintain high social distancing, i.e., significantly lower $\beta$. Yet it may be possible to lower $\beta$ early in the outbreak, and then go back to the normal $\beta$ later, when it is safe. The benefits are illustrated using simulations in figure \ref{Figure8}, where the chosen $\tau$, $\gamma$ and $p$ correspond to a critical $\beta = 1$
(recall equation~(\ref{stb1})). Now, suppose that normally $\beta = 1.04$. The disease will spread,
and saturate at
$S(\infty) \approx 0.92$, as per equation~(\ref{Sinf2}). Yet, for the same $\beta = 1.04$, a larger uninfected population of $S^*$ could be stable (equation~(\ref{Sstar})). That $S^*$, in principle,  {\em could be} reached using an artificially
low $\beta \approx 1.02$. If we change $\beta$ from 1.04 to 1.02 early in the outbreak, and hold $\beta = 1.02$ until a steady state is reached, then finally returning to $\beta = 1.04$ could yield a stable solution. The outbreak would be arrested, and the total number of affected people would be cut almost in half. 
In figure \ref{Figure8}, $\beta$ is switched from 1.04 to 1.02 at two different instants $T_c$, for two different simulations, and then switched back to 1.04 later. In each such switched case, although finally $\beta = 1.04$, the steady value
$S(\infty)$ corresponds to $\beta = 1.02$. In contrast, if $\beta = 1.04$ had been held throughout, almost twice as many people would have been infected.

\begin{figure}[htpb!]
	\begin{center}
		\includegraphics[width=0.48\textwidth]{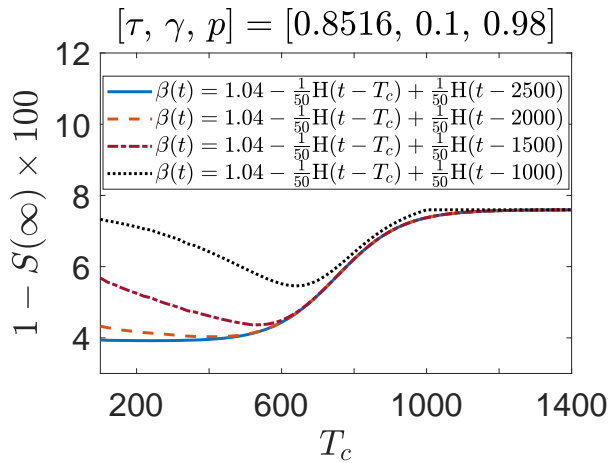}	
	\end{center}
	\caption{Influence of $T_c$ in $\beta(t)=1.04-0.02\textrm{H}(t-T_c)+0.02\textrm{H}(t-T_o)$ on the percentage of the total infected population at final stable steady state. The parameters used are the same as in figure~\ref{Figure8}, except for $T_c$ and $T_o$.}
	\label{Figure9}
\end{figure}

A further numerical study is presented in  figure  \ref{Figure9}. Here we vary both $T_c$ as well as $T_o$, the time of switching back to $\beta=1.04$. The parameters of figure  \ref{Figure8} are used except for $T_c$ and $T_o$. We see that the percentage of people affected can be cut almost in half for sufficiently low $T_c$ provided $T_o$ is large enough. If $T_o$ is made smaller, then there is a special value of $T_c$ when $S(\infty)$ is highest for the chosen $T_o$, but the gains may be suboptimal.

For higher $\beta$, more people must get infected (immune) before stability is achieved, and the benefit obtained is proportionately a little lower. For example, for the same $\tau$, $\gamma$ and $p$, if $\beta = 1.1$ is held fixed, then the final uninfected population is 82.4\%. In contrast, upon switching down to $\beta = 1.05$ for an extended period before returning to $\beta = 1.1$, the final uninfected population is 90.2\%, i.e., the number of infected people decreases by 46\%. The tradeoffs between $T_c$ and $T_o$ are similar to those observed in figure \ref{Figure9}, except that the dynamics is about two times faster (graphical results omitted for reasons of space).

The reader may anticipate that similar benefits can be obtained by lowering the time to quarantine, $\tau$, guided by the instability criterion of inequality (\ref{sat0}). However, with $p<1$ and $\gamma > 0$, the actual instability criterion is inequality (\ref{stb1}). Here, for example, if $p=0.9$ and $\gamma$ is tiny, then the stability condition is almost independent of $\tau$. However, if $p$ is close to unity and $\gamma$ is not too small (as in the parameters used in the foregoing examples), then the benefits of reducing $\tau$ can be significant. To demonstrate with a numerical example, the steady value $S(\infty)$ can be computed from equation~(\ref{longss})
using
$$p e^{-\gamma \tau} -1 + \frac{ 1 - p e^{-\gamma \tau} + \frac{\gamma \ln S(\infty)}{\beta}  }{S(\infty)} = 0.$$
With $\beta = 1$, $p=0.98$ and $\gamma = 0.1$, we obtain
table \ref{tab1}. It is seen that relatively small reductions in $\tau$ achieve significant reductions in the net population affected by the disease. However, if $p = 0.92$ (i.e., the chance of escaping quarantine is four times greater) and $\gamma$ remains the same, then instability occurs at $\tau = 0.22$ (about four times smaller), and $S(\infty)$ is less sensitive to $\tau$ as well (numerical examples omitted for brevity).
\begin{table}[htbp!]
	\caption{Role of $\tau$, with  $\beta = 1$, $p=0.98$ and $\gamma = 0.1$. \label{tab1}}
	\begin{center}
		\begin{tabular}{|l|l|l|l|}
			\hline
			$\tau$ & 0.8516 & 0.8716 & 0.8916 \\
			$S(\infty)$ & 1.000 & 0.965 & 0.931\\ \hline
		\end{tabular}
		
	\end{center}
\end{table}
We conclude that in circumstances where $p$ can be kept high (i.e., if public institutions are strong and detection followed by quarantine is nearly certain), and where the self-recovery rate $\gamma$ is not extremely small, the system can benefit significantly from even modest reductions in the detection time $\tau$. In such situations, research directed toward earlier detection may yield substantial benefits, and it may even be unnecessary to engage in extreme social distancing. On the other hand, if there is under-reporting and imperfect quarantining, then impractically large reductions in $\tau$ may be needed to compensate, and strict social distancing may be more effective.

\section{Concluding Discussion}
In this paper we have taken up a recently presented SEIQR model with delays. For a fast-spreading pandemic, loss of immunity of previously infected and cured people may reasonably be ignored. Under that simplification, the SEIQR model decouples so that only the $S$ and $I$ population equations need to be tackled.
It is known for this model that, for fixed parameter values in the unstable regime, an outbreak can occur. An initially small infected population can grow, and a significant portion of the original population can be affected.

We have first studied this model in some detail, seeking useful approximate solutions. For a weakly growing outbreak that affects a small proportion of the total population, and under a further simplification that neglects self-recovery and assumes perfect quarantining, the method of multiple scales yields an analytical expression for the complete progression of the outbreak, from infinitesimal initiation to final saturation. For moderate growth rates, a long wave approximation for the same parameters provides a nonlinear first order ODE for the same progression. With imperfect quarantining and nonzero self-recovery the long wave approximation for the full progression of the outbreak is given by a second order ODE. Finally, although the underlying DDE system is technically infinite dimensional, we have shown that a six-state Galerkin-based reduced order model for the system does an excellent job of capturing a wide range of solutions, i.e., the dynamics is effectively low-dimensional.

Subsequently, we have examined the implications of policy-induced social distancing, incorporated in our model as a time-varying infection rate $\beta(t)$. Interestingly and promisingly, we have found that an extended period of social distancing, imposed early in the outbreak, followed by an eventual relaxation to usual levels of interaction, can significantly lower the total numbers infected without losing stability of the final state. In the limit of weak growth, the number of infected people is cut in half. For faster growth, the reduction is a little smaller. Additionally, if the probability of an infected person being detected and quarantined is high, and the self-recovery rate not too small, then perhaps even stronger benefits can be obtained by slightly reducing the time until quarantine, $\tau$.

The above policy implications seem simple and robust. We emphasize that the benefit is not merely in lowering the rate at which people get infected, but also in the total number of people infected by the end of the outbreak. The intuitive key to understanding this reduction caused by social distancing lies in stability under fresh, but small, infection. Here, stability implies that with a small infected population, the outbreak will not grow very much
(recall figure \ref{Figure2a} versus \ref{Figure2b}). Under identical conditions, a larger infected population could cause the outbreak to grow: the assumption is that once the infected numbers are contained, a fresh large influx of infected people will be avoided. If $\beta$ is lowered with social distancing, the outbreak saturates at a high $S(\infty)$, and the infected population goes to near-zero values. Subsequently, under the assumption of no subsequent
large influx of infected people, $\beta$ can be increased within the stability boundary, and the outbreak does not grow significantly further.
In contrast, the benefits from reducing the time to quarantine, $\tau$, require greater sustained institutional alertness but may be stronger.

In closing, we must acknowledge that in any lumped model of the sort we study here, spatial variations in parameters and infected population densities are not modeled. Such lumped models are averaged models. Thus, it is not really clear at a detailed spatial level what it means to reduce the average infection rate $\beta$ by, say, 2\%. If the average person engages in social distancing, benefits will be seen on average, although there could still be localized outbreaks within pockets where people engage in riskier behavior. In constrast, however, the speed and probability of being detected and quarantined is less up to individual members of the public, and more in the hands of public institutions. Such institutions are amenable to tighter quality measures. So, from the viewpoint of reliability, we believe that large scale testing and near-certain isolation or quarantining can be critically useful in containing pandemics like COVID-19.

%\bibliographystyle{RS}
%\bibliography{DDEReferences}

\newpage
\section*{Appendix}
Here we outline our Galerkin projection calculation. Readers interested in the theoretical background may see, e.g., the so-called tau method of imposing boundary conditions in~\cite{Gottlieb1977}.

The initial functions for equations~(\ref{ge10}) and ~(\ref{ge11}) are assumed to be $S(t)=U_{1}(t),-\nu\le t\le0$ and $I(t)=U_{2}(t),-\nu\le t\le0$. Define $y_1(s,t)=S(t+s)$ and $y_2(s,t)=I(t+s)$. Equations~(\ref{ge10}) and~(\ref{ge11}) along with their history functions can be equivalently posed as the following partial differential equations with time dependent boundary conditions
\begin{eqnarray}
\label{app4}
\frac{\partial y_{1}}{\partial t}&=&\frac{\partial y_{1}}{\partial s},-\nu\le s\le0,\\
\label{app5}
\frac{\partial y_{2}}{\partial t}&=&\frac{\partial y_{2}}{\partial s},-\nu\le s\le0,\\
\label{app6}
\frac{\partial y_{1}}{\partial t}\bigg|_{s=0}&=&-\beta(t)y_{1}(0,t)y_{2}(0,t),\\
\label{app7}
\frac{\partial y_{2}}{\partial t}\bigg|_{s=0}&=&\beta(t-1) y_{1}(-1,t) y_{2}(-1,t)-\bar{p} \beta(t-\nu) y_{1}(-\nu,t) y_{2}(-\nu,t)-\gamma y_2(0,t).
\end{eqnarray}
Now, we assume a solution for $y_1(s,t)$ and $y_2(s,t)$ as follows
\begin{eqnarray}
\label{app8}
y_{1}(s,t)=\phi_1(s)\eta_1(t)+\phi_2(s)\eta_2(t)+\phi_3(s)\eta_3(t)\\
\label{app9}
y_{2}(s,t)=\phi_1(s)\eta_4(t)+\phi_2(s)\eta_5(t)+\phi_3(s)\eta_6(t)
\end{eqnarray}
The basis functions $\phi_1(s)=1$, $\phi_2(s)=1+\frac{2s}{\nu}$, and  $\phi_3(s)=\frac{3}{2}\left(1+\frac{2s}{\nu}\right)^{2}-\frac{1}{2}$ are shifted Legendre polynomials defined on the domain $-\nu\le s \le 0$. Substitute (\ref{app8}) into (\ref{app4}) and (\ref{app9}) into (\ref{app5}). Premultiplying each equation with $\phi_1(s)$ and then by $\phi_2(s)$ and integrating over the domain $-\nu\le s\le0$ each time,  we obtain
\begin{equation}
\label{app10}
\dot{\eta}_{1}=\frac{2}{\nu}\eta_{2}, \mbox{ and } \dot{\eta}_{2}=\frac{6}{\nu}\eta_{3}.
\end{equation}
\begin{equation}
\label{app11}
\dot{\eta}_{4}=\frac{2}{\nu}\eta_{5}, \mbox{ and } \dot{\eta}_{5}=\frac{6}{\nu}\eta_{6}. 
\end{equation}
The inner products with $\phi_3(s)$ are not taken. Instead, we substitute (\ref{app8}) and (\ref{app9}) in the boundary conditions  (\ref{app6}) and (\ref{app7}). There, we have $y_1(0,t)=\eta_1+\eta_2+\eta_3\,$, $y_2(0,t)=\eta_4+\eta_5+\eta_6\,$,  $y_1(-1,t)=\eta_1+\phi_2(-1)\eta_2+\phi_3(-1)\eta_3\,$,  $y_1(-\nu,t)=\eta_1-\eta_2+\eta_3\,$,  $y_2(-1,t)=\eta_4+\phi_2(-1)\eta_5+\phi_3(-1)\eta_6\,$,  and  $y_2(-\nu,t)=\eta_4-\eta_5+\eta_6$. Equations~(\ref{app6}) and ~(\ref{app7})  become
\begin{eqnarray}
\label{app12}
\dot{\eta}_{1}+\dot{\eta}_{2}+\dot{\eta}_{3}&=&-\beta(t)\left(\eta_1+\eta_2+\eta_3\right)\left(\eta_4+\eta_5+\eta_6\right),\\
\label{app13}
\dot{\eta}_{4}+\dot{\eta}_{5}+\dot{\eta}_{6}&=&\beta(t-1)\left(\eta_{1}+\alpha_{2}\eta_{2}+\alpha_{3}\eta_{3}\right)\left(\eta_{4}+\alpha_{2}\eta_{5}+\alpha_{3}\eta_{6}\right)\nonumber\\
&-&\bar{p}\beta(t-\nu)\left(\eta_{1}-\eta_{2}+\eta_{3}\right)\left(\eta_{4}-\eta_{5}+\eta_{6}\right)-\gamma\left(\eta_{4}+\eta_{5}+\eta_{6}\right),
\end{eqnarray}
giving us six ODEs for the six states, equivalent to equations~(\ref{eta1dot})-(\ref{eta6dot}).
The initial conditions for our ODEs can be obtained from history functions as $\eta_{k}(0)=\frac{2k-1}{\nu}\intop_{-\nu}^{0}U_{1}(s)\phi_{k}(s)ds,\,k=1,2,3$ and $\eta_{r}(0)=\frac{2(r-3)-1}{\nu}\intop_{-\nu}^{0}U_{2}(s)\phi_{r-3}(s)ds,\,r=4,5,6.$ 
\end{document}